\documentclass[pra,aps,showpacs,superscriptaddress,amsmath,amssymb]{revtex4}

\usepackage{graphicx}
\usepackage{dcolumn}
\usepackage{bm}
\usepackage{pstricks-add} 
\renewcommand{\|}{
\setlength{\unitlength}{3pt}
\psset{unit=3pt}
\psset{runit=2pt}
\psset{linewidth=0.2}
\begin{pspicture}(0,.5)(2.5,4)
\psline(1,0)(1,3)
\end{pspicture}}

\newcommand{\Y}{
\setlength{\unitlength}{3pt}
\psset{unit=3pt}
\psset{runit=2pt}
\psset{linewidth=0.2}
\begin{pspicture}(-1,.5)(3,4)
\psline(1,0)(1,2)
\psline(1,2)(0,3)
\psline(1,2)(2,3)
\end{pspicture}}

\newcommand\deuxun{
\setlength{\unitlength}{3pt}
\psset{unit=3pt}
\psset{runit=2pt}
\psset{linewidth=0.2}
\begin{pspicture}(0,.5)(5,5)
\psline(3,0)(3,2)
\psline(3,2)(1,4)
\psline(3,2)(4,3)
\psline(2,3)(3,4)
\end{pspicture}}

\newcommand\deuxdeux{
\setlength{\unitlength}{3pt}
\psset{unit=3pt}
\psset{runit=2pt}
\psset{linewidth=0.2}
\begin{pspicture}(1,.5)(6,5)
\psline(3,0)(3,2)
\psline(3,2)(5,4)
\psline(3,2)(2,3)
\psline(4,3)(3,4)
\end{pspicture}}

\newcommand\troisun{
\setlength{\unitlength}{3pt}
\psset{unit=3pt}
\psset{runit=2pt}
\psset{linewidth=0.2}
\begin{pspicture}(-1,.5)(5,6)
\psline(3,0)(3,2)
\psline(3,2)(0,5)
\psline(3,2)(4,3)
\psline(2,3)(3,4)
\psline(1,4)(2,5)
\end{pspicture}}

\newcommand\troisdeux{
\setlength{\unitlength}{3pt}
\psset{unit=3pt}
\psset{runit=2pt}
\psset{linewidth=0.2}
\begin{pspicture}(0,.5)(5,6)
\psline(3,0)(3,2)
\psline(3,2)(1,4)
\psline(3,2)(4,3)
\psline(2,3)(4,5)
\psline(3,4)(2,5)
\end{pspicture}}

\newcommand\troistrois{
\setlength{\unitlength}{3pt}
\psset{unit=3pt}
\psset{runit=2pt}
\psset{linewidth=0.2}
\begin{pspicture}(-0.5,.5)(6.5,6)
\psline(3,0)(3,2)
\psline(3,2)(0.5,4.5)
\psline(1.5,3.5)(2.5,4.5)
\psline(3,2)(5.5,4.5)
\psline(4.5,3.5)(3.5,4.5)
\end{pspicture}}

\newcommand\troisquatre{
\setlength{\unitlength}{3pt}
\psset{unit=3pt}
\psset{runit=2pt}
\psset{linewidth=0.2}
\begin{pspicture}(1,.5)(6,6)
\psline(3,0)(3,2)
\psline(3,2)(5,4)
\psline(3,2)(2,3)
\psline(4,3)(2,5)
\psline(3,4)(4,5)
\end{pspicture}}

\newcommand\troiscinq{
\setlength{\unitlength}{3pt}
\psset{unit=3pt}
\psset{runit=2pt}
\psset{linewidth=0.2}
\begin{pspicture}(1,.5)(7,6)
\psline(3,0)(3,2)
\psline(3,2)(6,5)
\psline(3,2)(2,3)
\psline(4,3)(3,4)
\psline(5,4)(4,5)
\end{pspicture}}

\newcommand\quatreun{
\setlength{\unitlength}{3pt}
\psset{unit=3pt}
\psset{runit=2pt}
\psset{linewidth=0.2}
\begin{pspicture}(3,.5)(5,6.5)
\psline(4,0)(4,2)
\psline(4,2)(0,6)
\psline(4,2)(5,3)
\psline(3,3)(4,4)
\psline(2,4)(3,5)
\psline(1,5)(2,6)
\end{pspicture}}

\newcommand\quatredeux{
\setlength{\unitlength}{3pt}
\psset{unit=3pt}
\psset{runit=2pt}
\psset{linewidth=0.2}
\begin{pspicture}(4,.5)(5,6.5)
\psline(4,0)(4,2)
\psline(4,2)(1,5)
\psline(4,2)(5,3)
\psline(3,3)(4,4)
\psline(2,4)(4,6)
\psline(3,5)(2,6)
\end{pspicture}}

\newcommand\quatretrois{
\setlength{\unitlength}{3pt}
\psset{unit=3pt}
\psset{runit=2pt}
\psset{linewidth=0.2}
\begin{pspicture}(4,.5)(6,6.5)
\psline(4,0)(4,2)
\psline(4,2)(0.5,5.5)
\psline(4,2)(5,3)
\psline(3,3)(5.5,5.5)
\psline(1.5,4.5)(2.5,5.5)
\psline(4.5,4.5)(3.5,5.5)
\end{pspicture}}

\newcommand\quatrequatre{
\setlength{\unitlength}{3pt}
\psset{unit=3pt}
\psset{runit=2pt}
\psset{linewidth=0.2}
\begin{pspicture}(3,.5)(6,6.5)
\psline(3,0)(3,2)
\psline(3,2)(1,4)
\psline(3,2)(4,3)
\psline(2,3)(4,5)
\psline(3,4)(1,6)
\psline(2,5)(3,6)
\end{pspicture}}

\newcommand\quatrecinq{
\setlength{\unitlength}{3pt}
\psset{unit=3pt}
\psset{runit=2pt}
\psset{linewidth=0.2}
\begin{pspicture}(3,.5)(5,6.5)
\psline(3,0)(3,2)
\psline(3,2)(1,4)
\psline(3,2)(4,3)
\psline(2,3)(5,6)
\psline(3,4)(2,5)
\psline(4,5)(3,6)
\end{pspicture}}

\newcommand\quatresix{
\setlength{\unitlength}{3pt}
\psset{unit=3pt}
\psset{runit=2pt}
\psset{linewidth=0.2}
\begin{pspicture}(3,.5)(6,6.5)
\psline(3.5,0)(3.5,2)
\psline(3.5,2)(0,5.5)
\psline(3.5,2)(6,4.5)
\psline(5,3.5)(4,4.5)
\psline(2,3.5)(3,4.5)
\psline(1,4.5)(2,5.5)
\end{pspicture}}

\newcommand\quatresept{
\setlength{\unitlength}{3pt}
\psset{unit=3pt}
\psset{runit=2pt}
\psset{linewidth=0.2}
\begin{pspicture}(2,.5)(5,6.5)
\psline(2.5,0)(2.5,2)
\psline(2.5,2)(0,4.5)
\psline(2.5,2)(5,4.5)
\psline(4,3.5)(3,4.5)
\psline(1,3.5)(3,5.5)
\psline(2,4.5)(1,5.5)
\end{pspicture}}

\newcommand\quatrehuit{
\setlength{\unitlength}{3pt}
\psset{unit=3pt}
\psset{runit=2pt}
\psset{linewidth=0.2}
\begin{pspicture}(2,.5)(5,6.5)
\psline(2.5,0)(2.5,2)
\psline(2.5,2)(5,4.5)
\psline(2.5,2)(0,4.5)
\psline(1,3.5)(2,4.5)
\psline(4,3.5)(2,5.5)
\psline(3,4.5)(4,5.5)
\end{pspicture}}

\newcommand\quatreneuf{
\setlength{\unitlength}{3pt}
\psset{unit=3pt}
\psset{runit=2pt}
\psset{linewidth=0.2}
\begin{pspicture}(3,.5)(6,6.5)
\psline(2.5,0)(2.5,2)
\psline(2.5,2)(6,5.5)
\psline(2.5,2)(0,4.5)
\psline(1,3.5)(2,4.5)
\psline(4,3.5)(3,4.5)
\psline(5,4.5)(4,5.5)
\end{pspicture}}

\newcommand\quatredix{
\setlength{\unitlength}{3pt}
\psset{unit=3pt}
\psset{runit=2pt}
\psset{linewidth=0.2}
\begin{pspicture}(3,.5)(5,6.5)
\psline(3,0)(3,2)
\psline(3,2)(5,4)
\psline(3,2)(2,3)
\psline(4,3)(1,6)
\psline(3,4)(4,5)
\psline(2,5)(3,6)
\end{pspicture}}

\newcommand\quatreonze{
\setlength{\unitlength}{3pt}
\psset{unit=3pt}
\psset{runit=2pt}
\psset{linewidth=0.2}
\begin{pspicture}(3,.5)(5,6.5)
\psline(2,0)(2,2)
\psline(2,2)(4,4)
\psline(2,2)(1,3)
\psline(3,3)(1,5)
\psline(2,4)(4,6)
\psline(3,5)(2,6)
\end{pspicture}}

\newcommand\quatredouze{
\setlength{\unitlength}{3pt}
\psset{unit=3pt}
\psset{runit=2pt}
\psset{linewidth=0.2}
\begin{pspicture}(3,.5)(6,6.5)
\psline(2,0)(2,2)
\psline(2,2)(5.5,5.5)
\psline(2,2)(1,3)
\psline(3,3)(0.5,5.5)
\psline(4.5,4.5)(3.5,5.5)
\psline(1.5,4.5)(2.5,5.5)
\end{pspicture}}

\newcommand\quatretreize{
\setlength{\unitlength}{3pt}
\psset{unit=3pt}
\psset{runit=2pt}
\psset{linewidth=0.2}
\begin{pspicture}(3,.5)(6,6.5)
\psline(2,0)(2,2)
\psline(2,2)(5,5)
\psline(2,2)(1,3)
\psline(3,3)(2,4)
\psline(4,4)(2,6)
\psline(3,5)(4,6)
\end{pspicture}}

\newcommand\quatrequatorze{
\setlength{\unitlength}{3pt}
\psset{unit=3pt}
\psset{runit=2pt}
\psset{linewidth=0.2}
\begin{pspicture}(3,.5)(6,6.5)
\psline(2,0)(2,2)
\psline(2,2)(6,6)
\psline(2,2)(1,3)
\psline(3,3)(2,4)
\psline(4,4)(3,5)
\psline(5,5)(4,6)
\end{pspicture}}
\newcommand{\lY}{
\setlength{\unitlength}{3pt}
\psset{unit=4pt}
\psset{runit=2pt}
\psset{linewidth=0.2}
\psset{labelsep=20}
\begin{pspicture}(2,.5)(3,4)
\rput[bl]{0}(1.4,1.3){$\scriptscriptstyle{1}$}
\psline(1,0)(1,2)
\psline(1,2)(0,3)
\psline(1,2)(2,3)
\end{pspicture}}

\newcommand\ldeuxun{
\setlength{\unitlength}{3pt}
\psset{unit=4pt}
\psset{runit=2pt}
\psset{linewidth=0.2}
\begin{pspicture}(2,.5)(5,5)
\rput[bl]{0}(3.4,1.3){$\scriptscriptstyle{1}$}
\rput[bl]{0}(1.0,2.3){$\scriptscriptstyle{2}$}
\psline(3,0)(3,2)
\psline(3,2)(1,4)
\psline(3,2)(4,3)
\psline(2,3)(3,4)
\end{pspicture}}

\newcommand\ldeuxdeux{
\setlength{\unitlength}{3pt}
\psset{unit=4pt}
\psset{runit=2pt}
\psset{linewidth=0.2}
\begin{pspicture}(2,.5)(6,5)
\rput[bl]{0}(3.4,1.3){$\scriptscriptstyle{1}$}
\rput[bl]{0}(4.5,2.3){$\scriptscriptstyle{2}$}
\psline(3,0)(3,2)
\psline(3,2)(5,4)
\psline(3,2)(2,3)
\psline(4,3)(3,4)
\end{pspicture}}

\newcommand\ltroisun{
\setlength{\unitlength}{3pt}
\psset{unit=4pt}
\psset{runit=2pt}
\psset{linewidth=0.2}
\begin{pspicture}(2,.5)(5,6)
\rput[bl]{0}(3.4,1.3){$\scriptscriptstyle{1}$}
\rput[bl]{0}(1.0,2.2){$\scriptscriptstyle{2}$}
\rput[bl]{0}(0.0,3.3){$\scriptscriptstyle{3}$}
\psline(3,0)(3,2)
\psline(3,2)(0,5)
\psline(3,2)(4,3)
\psline(2,3)(3,4)
\psline(1,4)(2,5)
\end{pspicture}}

\newcommand\ltroisdeux{
\setlength{\unitlength}{3pt}
\psset{unit=4pt}
\psset{runit=2pt}
\psset{linewidth=0.2}
\begin{pspicture}(2,.5)(5,6)
\rput[bl]{0}(3.4,1.3){$\scriptscriptstyle{1}$}
\rput[bl]{0}(1.0,2.2){$\scriptscriptstyle{2}$}
\rput[bl]{0}(3.4,3.5){$\scriptscriptstyle{3}$}
\psline(3,0)(3,2)
\psline(3,2)(1,4)
\psline(3,2)(4,3)
\psline(2,3)(4,5)
\psline(3,4)(2,5)
\end{pspicture}}

\newcommand\ltroistrois{
\setlength{\unitlength}{3pt}
\psset{unit=4pt}
\psset{runit=2pt}
\psset{linewidth=0.2}
\begin{pspicture}(2,.5)(6,6)
\rput[bl]{0}(3.4,1.3){$\scriptscriptstyle{1}$}
\rput[bl]{0}(4.6,2.4){$\scriptscriptstyle{2}$}
\rput[bl]{0}(0.9,2.4){$\scriptscriptstyle{3}$}
\psline(3,0)(3,2)
\psline(3,2)(0.5,4.5)
\psline(1.5,3.5)(2.5,4.5)
\psline(3,2)(5.5,4.5)
\psline(4.5,3.5)(3.5,4.5)
\end{pspicture}}

\newcommand\ltroisquatre{
\setlength{\unitlength}{3pt}
\psset{unit=4pt}
\psset{runit=2pt}
\psset{linewidth=0.2}
\begin{pspicture}(3,.5)(6,6)
\rput[bl]{0}(3.4,1.3){$\scriptscriptstyle{1}$}
\rput[bl]{0}(4.6,2.4){$\scriptscriptstyle{2}$}
\rput[bl]{0}(1.8,3.4){$\scriptscriptstyle{3}$}
\psline(3,0)(3,2)
\psline(3,2)(5,4)
\psline(3,2)(2,3)
\psline(4,3)(2,5)
\psline(3,4)(4,5)
\end{pspicture}}

\newcommand\ltroiscinq{
\setlength{\unitlength}{3pt}
\psset{unit=4pt}
\psset{runit=2pt}
\psset{linewidth=0.2}
\begin{pspicture}(3,.5)(6,6)
\rput[bl]{0}(3.4,1.3){$\scriptscriptstyle{1}$}
\rput[bl]{0}(4.4,2.3){$\scriptscriptstyle{2}$}
\rput[bl]{0}(5.4,3.3){$\scriptscriptstyle{3}$}
\psline(3,0)(3,2)
\psline(3,2)(6,5)
\psline(3,2)(2,3)
\psline(4,3)(3,4)
\psline(5,4)(4,5)
\end{pspicture}}

\newcommand\lquatreun{
\setlength{\unitlength}{3pt}
\psset{unit=4pt}
\psset{runit=2pt}
\psset{linewidth=0.2}
\begin{pspicture}(3,.5)(5,6.5)
\rput[bl]{0}(4.4,1.3){$\scriptscriptstyle{1}$}
\rput[bl]{0}(2.0,2.2){$\scriptscriptstyle{2}$}
\rput[bl]{0}(1.0,3.2){$\scriptscriptstyle{3}$}
\rput[bl]{0}(0.0,4.2){$\scriptscriptstyle{4}$}
\psline(4,0)(4,2)
\psline(4,2)(0,6)
\psline(4,2)(5,3)
\psline(3,3)(4,4)
\psline(2,4)(3,5)
\psline(1,5)(2,6)
\end{pspicture}}

\newcommand\lquatredeux{
\setlength{\unitlength}{3pt}
\psset{unit=4pt}
\psset{runit=2pt}
\psset{linewidth=0.2}
\begin{pspicture}(4,.5)(5,6.5)
\rput[bl]{0}(4.4,1.3){$\scriptscriptstyle{1}$}
\rput[bl]{0}(2.0,2.2){$\scriptscriptstyle{2}$}
\rput[bl]{0}(1.0,3.2){$\scriptscriptstyle{3}$}
\rput[bl]{0}(3.1,4.4){$\scriptscriptstyle{4}$}
\psline(4,0)(4,2)
\psline(4,2)(1,5)
\psline(4,2)(5,3)
\psline(3,3)(4,4)
\psline(2,4)(4,6)
\psline(3,5)(2,6)
\end{pspicture}}

\newcommand\lquatretrois{
\setlength{\unitlength}{3pt}
\psset{unit=4pt}
\psset{runit=2pt}
\psset{linewidth=0.2}
\begin{pspicture}(4,.5)(6,6.5)
\rput[bl]{0}(4.4,1.3){$\scriptscriptstyle{1}$}
\rput[bl]{0}(2.0,2.2){$\scriptscriptstyle{2}$}
\rput[bl]{0}(4.7,3.7){$\scriptscriptstyle{3}$}
\rput[bl]{0}(0.5,3.7){$\scriptscriptstyle{4}$}
\psline(4,0)(4,2)
\psline(4,2)(0.5,5.5)
\psline(4,2)(5,3)
\psline(3,3)(5.5,5.5)
\psline(1.5,4.5)(2.5,5.5)
\psline(4.5,4.5)(3.5,5.5)
\end{pspicture}}

\newcommand\lquatrequatre{
\setlength{\unitlength}{3pt}
\psset{unit=4pt}
\psset{runit=2pt}
\psset{linewidth=0.2}
\begin{pspicture}(3,.5)(6,6.5)
\rput[bl]{0}(3.4,1.3){$\scriptscriptstyle{1}$}
\rput[bl]{0}(1.0,2.2){$\scriptscriptstyle{2}$}
\rput[bl]{0}(3.1,3.2){$\scriptscriptstyle{3}$}
\rput[bl]{0}(1.1,4.2){$\scriptscriptstyle{4}$}
\psline(3,0)(3,2)
\psline(3,2)(1,4)
\psline(3,2)(4,3)
\psline(2,3)(4,5)
\psline(3,4)(1,6)
\psline(2,5)(3,6)
\end{pspicture}}

\newcommand\lquatrecinq{
\setlength{\unitlength}{3pt}
\psset{unit=4pt}
\psset{runit=2pt}
\psset{linewidth=0.2}
\begin{pspicture}(3,.5)(5,6.5)
\rput[bl]{0}(3.4,1.3){$\scriptscriptstyle{1}$}
\rput[bl]{0}(1.0,2.2){$\scriptscriptstyle{2}$}
\rput[bl]{0}(3.1,3.2){$\scriptscriptstyle{3}$}
\rput[bl]{0}(4.0,4.2){$\scriptscriptstyle{4}$}
\psline(3,0)(3,2)
\psline(3,2)(1,4)
\psline(3,2)(4,3)
\psline(2,3)(5,6)
\psline(3,4)(2,5)
\psline(4,5)(3,6)
\end{pspicture}}

\newcommand\lquatresix{
\setlength{\unitlength}{3pt}
\psset{unit=4pt}
\psset{runit=2pt}
\psset{linewidth=0.2}
\begin{pspicture}(3,.5)(6,6.5)
\rput[bl]{0}(3.9,1.3){$\scriptscriptstyle{1}$}
\rput[bl]{0}(5.2,2.7){$\scriptscriptstyle{2}$}
\rput[bl]{0}(1.2,2.7){$\scriptscriptstyle{3}$}
\rput[bl]{0}(0.2,3.7){$\scriptscriptstyle{4}$}
\psline(3.5,0)(3.5,2)
\psline(3.5,2)(0,5.5)
\psline(3.5,2)(6,4.5)
\psline(5,3.5)(4,4.5)
\psline(2,3.5)(3,4.5)
\psline(1,4.5)(2,5.5)
\end{pspicture}}

\newcommand\lquatresept{
\setlength{\unitlength}{3pt}
\psset{unit=4pt}
\psset{runit=2pt}
\psset{linewidth=0.2}
\begin{pspicture}(2,.5)(5,6.5)
\rput[bl]{0}(2.9,1.3){$\scriptscriptstyle{1}$}
\rput[bl]{0}(4.2,2.7){$\scriptscriptstyle{2}$}
\rput[bl]{0}(0.2,2.7){$\scriptscriptstyle{3}$}
\rput[bl]{0}(1.8,3.7){$\scriptscriptstyle{4}$}
\psline(2.5,0)(2.5,2)
\psline(2.5,2)(0,4.5)
\psline(2.5,2)(5,4.5)
\psline(4,3.5)(3,4.5)
\psline(1,3.5)(3,5.5)
\psline(2,4.5)(1,5.5)
\end{pspicture}}

\newcommand\lquatrehuit{
\setlength{\unitlength}{3pt}
\psset{unit=4pt}
\psset{runit=2pt}
\psset{linewidth=0.2}
\begin{pspicture}(2,.5)(5,6.5)
\rput[bl]{0}(2.9,1.3){$\scriptscriptstyle{1}$}
\rput[bl]{0}(4.2,2.7){$\scriptscriptstyle{2}$}
\rput[bl]{0}(2.2,3.7){$\scriptscriptstyle{3}$}
\rput[bl]{0}(0.2,2.7){$\scriptscriptstyle{4}$}
\psline(2.5,0)(2.5,2)
\psline(2.5,2)(5,4.5)
\psline(2.5,2)(0,4.5)
\psline(1,3.5)(2,4.5)
\psline(4,3.5)(2,5.5)
\psline(3,4.5)(4,5.5)
\end{pspicture}}

\newcommand\lquatreneuf{
\setlength{\unitlength}{3pt}
\psset{unit=4pt}
\psset{runit=2pt}
\psset{linewidth=0.2}
\begin{pspicture}(3,.5)(6,6.5)
\rput[bl]{0}(2.9,1.3){$\scriptscriptstyle{1}$}
\rput[bl]{0}(4.2,2.7){$\scriptscriptstyle{2}$}
\rput[bl]{0}(5.2,3.7){$\scriptscriptstyle{3}$}
\rput[bl]{0}(0.2,2.7){$\scriptscriptstyle{4}$}
\psline(2.5,0)(2.5,2)
\psline(2.5,2)(6,5.5)
\psline(2.5,2)(0,4.5)
\psline(1,3.5)(2,4.5)
\psline(4,3.5)(3,4.5)
\psline(5,4.5)(4,5.5)
\end{pspicture}}

\newcommand\lquatredix{
\setlength{\unitlength}{3pt}
\psset{unit=4pt}
\psset{runit=2pt}
\psset{linewidth=0.2}
\begin{pspicture}(3,.5)(5,6.5)
\rput[bl]{0}(3.4,1.3){$\scriptscriptstyle{1}$}
\rput[bl]{0}(4.4,2.3){$\scriptscriptstyle{2}$}
\rput[bl]{0}(2.1,3.2){$\scriptscriptstyle{3}$}
\rput[bl]{0}(1.1,4.2){$\scriptscriptstyle{4}$}
\psline(3,0)(3,2)
\psline(3,2)(5,4)
\psline(3,2)(2,3)
\psline(4,3)(1,6)
\psline(3,4)(4,5)
\psline(2,5)(3,6)
\end{pspicture}}

\newcommand\lquatreonze{
\setlength{\unitlength}{3pt}
\psset{unit=4pt}
\psset{runit=2pt}
\psset{linewidth=0.2}
\begin{pspicture}(3,.5)(5,6.5)
\rput[bl]{0}(2.4,1.3){$\scriptscriptstyle{1}$}
\rput[bl]{0}(3.4,2.3){$\scriptscriptstyle{2}$}
\rput[bl]{0}(1.1,3.2){$\scriptscriptstyle{3}$}
\rput[bl]{0}(3.1,4.3){$\scriptscriptstyle{4}$}
\psline(2,0)(2,2)
\psline(2,2)(4,4)
\psline(2,2)(1,3)
\psline(3,3)(1,5)
\psline(2,4)(4,6)
\psline(3,5)(2,6)
\end{pspicture}}

\newcommand\lquatredouze{
\setlength{\unitlength}{3pt}
\psset{unit=4pt}
\psset{runit=2pt}
\psset{linewidth=0.2}
\begin{pspicture}(3,.5)(6,6.5)
\rput[bl]{0}(2.4,1.3){$\scriptscriptstyle{1}$}
\rput[bl]{0}(3.4,2.3){$\scriptscriptstyle{2}$}
\rput[bl]{0}(4.7,3.7){$\scriptscriptstyle{3}$}
\rput[bl]{0}(0.5,3.7){$\scriptscriptstyle{4}$}
\psline(2,0)(2,2)
\psline(2,2)(5.5,5.5)
\psline(2,2)(1,3)
\psline(3,3)(0.5,5.5)
\psline(4.5,4.5)(3.5,5.5)
\psline(1.5,4.5)(2.5,5.5)
\end{pspicture}}

\newcommand\lquatretreize{
\setlength{\unitlength}{3pt}
\psset{unit=4pt}
\psset{runit=2pt}
\psset{linewidth=0.2}
\begin{pspicture}(3,.5)(6,6.5)
\rput[bl]{0}(2.4,1.3){$\scriptscriptstyle{1}$}
\rput[bl]{0}(3.4,2.3){$\scriptscriptstyle{2}$}
\rput[bl]{0}(4.4,3.3){$\scriptscriptstyle{3}$}
\rput[bl]{0}(2.1,4.3){$\scriptscriptstyle{4}$}
\psline(2,0)(2,2)
\psline(2,2)(5,5)
\psline(2,2)(1,3)
\psline(3,3)(2,4)
\psline(4,4)(2,6)
\psline(3,5)(4,6)
\end{pspicture}}

\newcommand\lquatrequatorze{
\setlength{\unitlength}{3pt}
\psset{unit=4pt}
\psset{runit=2pt}
\psset{linewidth=0.2}
\begin{pspicture}(3,.5)(6,6.5)
\rput[bl]{0}(2.4,1.3){$\scriptscriptstyle{1}$}
\rput[bl]{0}(3.4,2.3){$\scriptscriptstyle{2}$}
\rput[bl]{0}(4.4,3.3){$\scriptscriptstyle{3}$}
\rput[bl]{0}(5.4,4.3){$\scriptscriptstyle{4}$}
\psline(2,0)(2,2)
\psline(2,2)(6,6)
\psline(2,2)(1,3)
\psline(3,3)(2,4)
\psline(4,4)(3,5)
\psline(5,5)(4,6)
\end{pspicture}}

\newcommand{\DY}{
\setlength{\unitlength}{5pt}
\psset{unit=5pt}
\psset{runit=2pt}
\psset{linewidth=0.12}
\begin{pspicture}(2,0)(2,2)
\psdots[dotsize=1.5pt](0,0)(1,1)(2,0)
\psline(0,0)(1,1)
\psline(1,1)(2,0)
\psline(0,0)(2,0)
\end{pspicture}}

\newcommand\Ddeuxun{
\setlength{\unitlength}{3pt}
\psset{unit=3pt}
\psset{runit=2pt}
\psset{linewidth=0.2}
\begin{pspicture}(3,0)(4,2)
\psdots[dotsize=1.5pt](0,0)(1,1)(2,2)(3,1)(4,0)
\psline(0,0)(2,2)
\psline(2,2)(4,0)
\psline(0,0)(4,0)
\end{pspicture}}

\newcommand\Ddeuxdeux{
\setlength{\unitlength}{3pt}
\psset{unit=3pt}
\psset{runit=2pt}
\psset{linewidth=0.2}
\begin{pspicture}(3,0)(4,2)
\psdots[dotsize=1.5pt](0,0)(1,1)(2,0)(3,1)(4,0)
\psline(0,0)(1,1)
\psline(1,1)(2,0)
\psline(2,0)(3,1)
\psline(3,1)(4,0)
\psline(0,0)(4,0)
\end{pspicture}}

\newcommand\Dtroisun{
\setlength{\unitlength}{3pt}
\psset{unit=3pt}
\psset{runit=2pt}
\psset{linewidth=0.2}
\begin{pspicture}(3,0)(6,3)
\psdots[dotsize=1.5pt](0,0)(1,1)(2,2)(3,3)(4,2)(5,1)(6,0)
\psline(0,0)(3,3)
\psline(3,3)(6,0)
\psline(0,0)(6,0)
\end{pspicture}}

\newcommand\Dtroisdeux{
\setlength{\unitlength}{3pt}
\psset{unit=3pt}
\psset{runit=2pt}
\psset{linewidth=0.2}
\begin{pspicture}(3,0)(6,3)
\psdots[dotsize=1.5pt](0,0)(1,1)(2,2)(3,1)(4,2)(5,1)(6,0)
\psline(0,0)(2,2)
\psline(2,2)(3,1)
\psline(3,1)(4,2)
\psline(4,2)(6,0)
\psline(0,0)(6,0)
\end{pspicture}}

\newcommand\Dtroistrois{
\setlength{\unitlength}{3pt}
\psset{unit=3pt}
\psset{runit=2pt}
\psset{linewidth=0.2}
\begin{pspicture}(3,0)(6,2)
\psdots[dotsize=1.5pt](0,0)(1,1)(2,2)(3,1)(4,0)(5,1)(6,0)
\psline(0,0)(2,2)
\psline(2,2)(3,1)
\psline(3,1)(4,0)
\psline(4,0)(5,1)
\psline(5,1)(6,0)
\psline(0,0)(6,0)
\end{pspicture}}

\newcommand\Dtroisquatre{
\setlength{\unitlength}{3pt}
\psset{unit=3pt}
\psset{runit=2pt}
\psset{linewidth=0.2}
\begin{pspicture}(3,0)(6,2)
\psdots[dotsize=1.5pt](0,0)(1,1)(2,0)(3,1)(4,2)(5,1)(6,0)
\psline(0,0)(1,1)
\psline(1,1)(2,0)
\psline(2,0)(4,2)
\psline(4,2)(6,0)
\psline(0,0)(6,0)
\end{pspicture}}

\newcommand\Dtroiscinq{
\setlength{\unitlength}{3pt}
\psset{unit=3pt}
\psset{runit=2pt}
\psset{linewidth=0.2}
\begin{pspicture}(3,0)(6,1)
\psdots[dotsize=1.5pt](0,0)(1,1)(2,0)(3,1)(4,0)(5,1)(6,0)
\psline(0,0)(1,1)
\psline(1,1)(2,0)
\psline(2,0)(3,1)
\psline(3,1)(4,0)
\psline(4,0)(5,1)
\psline(5,1)(6,0)
\psline(0,0)(6,0)
\end{pspicture}}

\newcommand\Dquatreun{
\setlength{\unitlength}{3pt}
\psset{unit=3pt}
\psset{runit=2pt}
\psset{linewidth=0.2}
\begin{pspicture}(4,0)(8,4)
\psdots[dotsize=1.5pt](0,0)(1,1)(2,2)(3,3)(4,4)(5,3)(6,2)(7,1)(8,0)
\psline(0,0)(4,4)
\psline(4,4)(8,0)
\psline(0,0)(8,0)
\end{pspicture}}

\newcommand\Dquatredeux{
\setlength{\unitlength}{3pt}
\psset{unit=3pt}
\psset{runit=2pt}
\psset{linewidth=0.2}
\begin{pspicture}(4,0)(8,3)
\psdots[dotsize=1.5pt](0,0)(1,1)(2,2)(3,3)(4,2)(5,3)(6,2)(7,1)(8,0)
\psline(0,0)(3,3)
\psline(3,3)(4,2)
\psline(4,2)(5,3)
\psline(5,3)(8,0)
\psline(0,0)(8,0)
\end{pspicture}}

\newcommand\Dquatretrois{
\setlength{\unitlength}{3pt}
\psset{unit=3pt}
\psset{runit=2pt}
\psset{linewidth=0.2}
\begin{pspicture}(4,0)(8,3)
\psdots[dotsize=1.5pt](0,0)(1,1)(2,2)(3,3)(4,2)(5,1)(6,2)(7,1)(8,0)
\psline(0,0)(3,3)
\psline(3,3)(5,1)
\psline(5,1)(6,2)
\psline(6,2)(8,0)
\psline(0,0)(8,0)
\end{pspicture}}

\newcommand\Dquatrequatre{
\setlength{\unitlength}{3pt}
\psset{unit=3pt}
\psset{runit=2pt}
\psset{linewidth=0.2}
\begin{pspicture}(4,0)(8,3)
\psdots[dotsize=1.5pt](0,0)(1,1)(2,2)(3,1)(4,2)(5,3)(6,2)(7,1)(8,0)
\psline(0,0)(2,2)
\psline(2,2)(3,1)
\psline(3,1)(5,3)
\psline(5,3)(8,0)
\psline(0,0)(8,0)
\end{pspicture}}

\newcommand\Dquatrecinq{
\setlength{\unitlength}{3pt}
\psset{unit=3pt}
\psset{runit=2pt}
\psset{linewidth=0.2}
\begin{pspicture}(4,0)(8,3)
\psdots[dotsize=1.5pt](0,0)(1,1)(2,2)(3,1)(4,2)(5,1)(6,2)(7,1)(8,0)
\psline(0,0)(2,2)
\psline(2,2)(3,1)
\psline(3,1)(4,2)
\psline(4,2)(5,1)
\psline(5,1)(6,2)
\psline(6,2)(8,0)
\psline(0,0)(8,0)
\end{pspicture}}

\newcommand\Dquatresix{
\setlength{\unitlength}{3pt}
\psset{unit=3pt}
\psset{runit=2pt}
\psset{linewidth=0.2}
\begin{pspicture}(4,0)(8,3)
\psdots[dotsize=1.5pt](0,0)(1,1)(2,2)(3,3)(4,2)(5,1)(6,0)(7,1)(8,0)
\psline(0,0)(3,3)
\psline(3,3)(6,0)
\psline(6,0)(7,1)
\psline(7,1)(8,0)
\psline(0,0)(8,0)
\end{pspicture}}

\newcommand\Dquatresept{
\setlength{\unitlength}{3pt}
\psset{unit=3pt}
\psset{runit=2pt}
\psset{linewidth=0.2}
\begin{pspicture}(4,0)(8,2)
\psdots[dotsize=1.5pt](0,0)(1,1)(2,2)(3,1)(4,2)(5,1)(6,0)(7,1)(8,0)
\psline(0,0)(2,2)
\psline(2,2)(3,1)
\psline(3,1)(4,2)
\psline(4,2)(6,0)
\psline(6,0)(7,1)
\psline(7,1)(8,0)
\psline(0,0)(8,0)
\end{pspicture}}

\newcommand\Dquatrehuit{
\setlength{\unitlength}{3pt}
\psset{unit=3pt}
\psset{runit=2pt}
\psset{linewidth=0.2}
\begin{pspicture}(4,0)(8,2)
\psdots[dotsize=1.5pt](0,0)(1,1)(2,2)(3,1)(4,0)(5,1)(6,2)(7,1)(8,0)
\psline(0,0)(2,2)
\psline(2,2)(4,0)
\psline(4,0)(6,2)
\psline(6,2)(8,0)
\psline(0,0)(8,0)
\end{pspicture}}

\newcommand\Dquatreneuf{
\setlength{\unitlength}{3pt}
\psset{unit=3pt}
\psset{runit=2pt}
\psset{linewidth=0.2}
\begin{pspicture}(4,0)(8,2)
\psdots[dotsize=1.5pt](0,0)(1,1)(2,2)(3,1)(4,0)(5,1)(6,0)(7,1)(8,0)
\psline(0,0)(2,2)
\psline(2,2)(4,0)
\psline(4,0)(5,1)
\psline(5,1)(6,0)
\psline(6,0)(7,1)
\psline(7,1)(8,0)
\psline(0,0)(8,0)
\end{pspicture}}

\newcommand\Dquatredix{
\setlength{\unitlength}{3pt}
\psset{unit=3pt}
\psset{runit=2pt}
\psset{linewidth=0.2}
\begin{pspicture}(4,0)(8,3)
\psdots[dotsize=1.5pt](0,0)(1,1)(2,0)(3,1)(4,2)(5,3)(6,2)(7,1)(8,0)
\psline(0,0)(1,1)
\psline(1,1)(2,0)
\psline(2,0)(5,3)
\psline(5,3)(8,0)
\psline(0,0)(8,0)
\end{pspicture}}

\newcommand\Dquatreonze{
\setlength{\unitlength}{3pt}
\psset{unit=3pt}
\psset{runit=2pt}
\psset{linewidth=0.2}
\begin{pspicture}(4,0)(8,2)
\psdots[dotsize=1.5pt](0,0)(1,1)(2,0)(3,1)(4,2)(5,1)(6,2)(7,1)(8,0)
\psline(0,0)(1,1)
\psline(1,1)(2,0)
\psline(2,0)(4,2)
\psline(4,2)(5,1)
\psline(5,1)(6,2)
\psline(6,2)(8,0)
\psline(0,0)(8,0)
\end{pspicture}}

\newcommand\Dquatredouze{
\setlength{\unitlength}{3pt}
\psset{unit=3pt}
\psset{runit=2pt}
\psset{linewidth=0.2}
\begin{pspicture}(4,0)(8,2)
\psdots[dotsize=1.5pt](0,0)(1,1)(2,0)(3,1)(4,2)(5,1)(6,0)(7,1)(8,0)
\psline(0,0)(1,1)
\psline(1,1)(2,0)
\psline(2,0)(4,2)
\psline(4,2)(6,0)
\psline(6,0)(7,1)
\psline(7,1)(8,0)
\psline(0,0)(8,0)
\end{pspicture}}

\newcommand\Dquatretreize{
\setlength{\unitlength}{3pt}
\psset{unit=3pt}
\psset{runit=2pt}
\psset{linewidth=0.2}
\begin{pspicture}(4,0)(8,2)
\psdots[dotsize=1.5pt](0,0)(1,1)(2,0)(3,1)(4,0)(5,1)(6,2)(7,1)(8,0)
\psline(0,0)(1,1)
\psline(1,1)(2,0)
\psline(2,0)(3,1)
\psline(3,1)(4,0)
\psline(4,0)(6,2)
\psline(6,2)(8,0)
\psline(0,0)(8,0)
\end{pspicture}}

\newcommand\Dquatrequatorze{
\setlength{\unitlength}{3pt}
\psset{unit=3pt}
\psset{runit=2pt}
\psset{linewidth=0.2}
\begin{pspicture}(4,0)(8,1)
\psdots[dotsize=1.5pt](0,0)(1,1)(2,0)(3,1)(4,0)(5,1)(6,0)(7,1)(8,0)
\psline(0,0)(1,1)
\psline(1,1)(2,0)
\psline(2,0)(3,1)
\psline(3,1)(4,0)
\psline(4,0)(5,1)
\psline(5,1)(6,0)
\psline(6,0)(7,1)
\psline(7,1)(8,0)
\psline(0,0)(8,0)
\end{pspicture}}

\newcommand{\hatOmega}{\hat{\Omega}}
\newcommand{\calL}{\mathcal{L}}

\newcommand{\Heff}{{H_{\mathrm{eff}}}}
\newcommand{\Pp}[1]{{\overline{P}_{#1}}}
\newcommand{\brap}[1]{{|\overline{#1}\rangle}}
\newcommand{\ketp}[1]{{\langle\overline{#1}|}}
\newcommand{\enep}[1]{{\overline{e}_{#1}}}

\newcommand{\lrgraft}[2]
{\setlength{\unitlength}{3pt}
\psset{unit=3pt}
\psset{runit=2pt}
\psset{linewidth=0.2}
\begin{pspicture}(0,0)(6,3.5)
\psline(3,-1)(3,.5)
\psline(3,.5)(2,1.5)
\psline(3,.5)(4,1.5)
\put(0.5,3.5){$#1$}
\put(6.5,3.5){$#2$}
\end{pspicture}}

\begin{document}

\preprint{APS/123-QED}

\title{The Rayleigh-Schr{\"o}dinger perturbation series of quasi-degenerate systems}

\author{Christian Brouder}
\affiliation{%
Institut de Min\'eralogie et de Physique des Milieux
Condens\'es, \\
Universit\'e Pierre et Marie Curie-Paris 6,
CNRS UMR7590,
\\
Universit\'e Denis Diderot-Paris 7,
Institut de Physique du Globe,\\
Campus Jussieu, bo{\^{\i}}te courrier 115,
4 place Jussieu, 75252 Paris cedex 05, France.
}%
\author{G\'erard H. E. Duchamp}
\affiliation{Laboratoire d'Informatique de Paris-Nord, CNRS UMR 7030,
Institut Galil\'ee, Universit\'e Paris-Nord,
99 avenue Jean-Baptiste Cl\'ement, 93430 Villetaneuse.
}%
\author{Fr\'ed\'eric Patras}
\affiliation{Laboratoire J.-A. Dieudonn\'e, CNRS UMR 6621,
Universit\'e de Nice,
Parc Valrose, 06108 Nice Cedex 02, France.
}%
\author{G. Z. T{\'o}th}
\affiliation{Theoretical Department,
Research Institute for Particle and
  Nuclear Physics, P.O. box 49, Budapest, 1525, Hungary.
}%

\date{\today}

\begin{abstract}
We present the first representation of the general term of the 
Rayleigh-Schr\"odinger series for quasidegenerate systems. 
Each term of the series is represented by a tree and there is a
straightforward relation between the tree and the analytical
expression of the corresponding term.
The combinatorial and graphical techniques used in the proof of 
the series expansion allow us to derive various resummation formulas 
of the series. The relation with several combinatorial objects
used for special cases (degenerate or non-degenerate systems)
is established.
\end{abstract}

\maketitle

\section{Introduction}

Rayleigh-Schr\"odinger (RS) perturbation theory is a venerable
technique to calculate the eigenvalues and eigenvectors
of $H=H_0+V$ from those of $H_0$.
It was created in 1894 by Lord Rayleigh 
to describe the vibrations of a string~\cite{Rayleigh-94} and adapted
to quantum mechanics by Schr\"odinger in 1926~\cite{Schrodinger-26}.
RS perturbation theory has been used in all fields of quantum
physics (particle, atomic, molecular, solid-state physics)
and quantum chemistry.

In most textbooks, the RS series deals with the perturbation of a single
nondegenerate initial state. 
However, in many practical applications, the initial
state is either degenerate or quasidegenerate (i.e. several states
are close in energy) and the basic RS approach breaks down.
Quasidegenerate perturbation theory is 
widely used to set up effective Hamiltonians~\cite{LindgrenMorrison},
for example the spin Hamiltonians of 
molecular chemistry and solid state physics~\cite{Moreira,Moreira-02},
or to deal with the quantum electrodynamics of 
atoms~\cite{Lindgren1,Lebigot}.

The terms of the RS series are notoriously complex.
Even mathematical physicists of the stature of Reed and Simon
admit that the terms of the RS series are ``quite complicated''
and that ``the higher order RS coefficients are hard to 
compute'' (see ref.~\onlinecite{ReedSimonIV}, p.~8 and 18).
Computer programs are available to build the terms of the
RS series~\cite{Barnett-04,Hirata-06,Fritzsche-08,Derevianko-10,Jursenas-10},
but their results are intricate expressions that do not exhibit
any obvious structure and that can hardly
be used to carry out resummations of the series.

In this paper, combinatorial physics is used to provide the first
representation of the general term of the Rayleigh-Schr\"odinger (RS)
perturbation theory for quasidegenerate systems.
Each term is written as a tree that faithfully reflects its
algebraic structure. In particular, trees illustrate
the recursive structure of RS terms, that is 
used to prove properties of the RS series.

The purpose of this paper is to describe the general term of the RS
series and to illustrate the power of our combinatorial approach
by deriving a number of possible resummations of the RS series.
Our aim is to set up the tools for performing such resummations and
not to discuss when some resummations are more convenient
than others. As a consequence, no numerical example is given.

When the system is not quasi-degenerate (i.e. when it is
either fully degenerate or non-degenerate), several graphical
representations of the terms of the RS series have been
proposed. We give the bijection between these representations
and our trees.

\section{Rayleigh-Schr\"odinger series}
In the most general setting, we consider a model space $M$ spanned by
$N$ eigenstates $|i\rangle$ of $H_0$, with energies $e_i$.
We assume that all $e_i$ of the model space 
are separated from the rest of the spectrum of $H_0$
by a finite gap. The projector onto the model space $M$ is
$P=\sum_i |i\rangle\langle i|$, where $i$ runs over the basis states of $M$.
The wave operator $\Omega$ transforms $N$ states $|\phi_i\rangle$ of $M$
into eigenstates $|\Phi_i\rangle=\Omega|\phi_i\rangle$ of $H$: 
$H\Omega|\phi_i\rangle=E_i\Omega|\phi_i\rangle$.
We recently described~\cite{BMP} the way to choose
the states $|\phi_i\rangle$.

We assume that $P\Omega=P$ and $\Omega P=\Omega$.
Then, the eigenvalues of the effective Hamiltonian $\Heff=PH\Omega$ 
are the eigenvalues $E_i$ of $H$. In other words, eigenvalues of $H$
can be obtained by diagonalizing the $M\times M$ matrix $\Heff$.
Lindgren and Kvasni{\v{c}}ka showed~\cite{Lindgren74,Kvasnicka-74} that
\begin{eqnarray}
{[}\Omega,H_0{]} &=& V\Omega - \Omega V \Omega.
\label{Lindgren}
\end{eqnarray}
The recursive solution of this equation gives a series expansion
for $\Omega$ which is the RS series for the wavefunction.
However, it is not obvious that eq.~(\ref{Lindgren}) has
a recursive solution. Indeed, we must be able to solve the
equation ${[}X,H_0{]} = C$ for $X$.
In general, the operator Sylvester equation
$AX-XB=C$, where $A$ and $B$ are self-adjoint, has
a unique solution if and only if $A$ and $B$ have no
common eigenvalue~\cite{Rutherford-32,Albeverio-09}. Clearly, this is not
the case if $A=B=H_0$, so we recast the equation
by defining $\chi=\Omega-P$, where $\chi=Q\chi P$,
with $Q=1-P$. Thus, eq.~(\ref{Lindgren})
becomes an equation for $\chi$:
\begin{eqnarray}
{[}\chi,H_0{]} &=& QVP +QV\chi -\chi V P - \chi V \chi.
\label{Lindgrenchi}
\end{eqnarray}
This equation is a matrix Riccati equation,
for which various solution methods have been 
proposed in the 
literature~\cite{Nair-01,Kostrykin-05,Albeverio-09,Fujii-10}.
Solutions exist also for its time-dependent 
form~\cite{Common-86,Common-90}.
Now, 
${[}\chi,H_0{]} = \chi H_0 - H_0 \chi
= \chi P H_0 - Q H_0 \chi$,
because $P$ and $Q$ commute with $H_0$.
We obtain a Sylvester equation with
$A=QH_0$ and $B=P H_0$ and its solution is unique 
because we assumed that
there is a finite gap between the states of the model
space and the rest of the spectrum.

\subsection{Combinatorial analysis}
Because of its importance, the basic RS series has been dealt with in
hundreds of papers.
As a starting point to discover the proper
combinatorial structure of the RS series, we enumerate the number
of its terms: it has one term of order 0, one term of order 1,
then 2, 5, 14, 42, 132, 429 terms of order
2, 3, 4, 5, 6 and 7, respectively. The number of terms of
order $n$ is the Catalan number
$C_n=\frac{(2n)!}{n!(n+1)!}$.
The Catalan sequence is well-know in combinatorics:
``The Catalan sequence is probably the most frequently encountered
sequence that is still obscure enough to cause mathematicians
[\dots] to expend inordinate amounts of energy re-discovering
formulas that were worked out long ago.''~\cite{Gardner-76}.
An entire book is devoted to Catalan numbers~\cite{Koshy}.
They enumerate over
200 combinatorial objects~\cite{StanleyI} and 
in the physics literature, several of these objects were used to represent the
general term of the RS series: 
Dyck paths~\cite{Bloch-58}, non-crossing partitions~\cite{Olszewski-04},
bracketings~\cite{Huby-61,Tong-62,Lindgren74}, 
labelled diagrams~\cite{Salzman-68}
and
sequences of integers~\cite{Bloch-58,Silverstone-71,Suzuki-84}.
However, these approaches could not be generalized to
quasi-degenerate systems because they dealt with 
denominators $(e_0-e_j)^n$ with $n>1$, whereas
more complex denominators occur in the quasi-degenerate case.
Moreover, they could not be used
to discover efficient resummations because they
do not capture the recursive structure
of the RS series. 

To find out which combinatorial object is best suited to deal with
the RS series, we had to use a rather elaborate algebraic analysis
of the time-dependent perturbation theory~\cite{BP-09,BMP}.
It turned out in the end that 
planar binary trees provide the most faithful representation.
As we shall see, the relation beween a tree and the corresponding term of RS series
is straightforward and the recursive structure of the
terms is transparent. 
The faithfulness is demonstrated
by the easiness of the proofs and by the relation
between geometrical properties of the trees and
analytical properties of the terms, for instance
for resummation.

\subsection{Binary trees}
We first give examples of the planar binary trees that we use.
We denote by $Y_n$ the set of planar binary trees with $n$
inner vertices.
A tree is called \emph{binary} if each vertex has either zero
or two children. It is called \emph{planar} 
if two trees are different when they can be deduced one from the other by
moving one edge over another one. For example, the two
planar trees $\deuxun$ and $\deuxdeux$ are different.

There is a single tree with zero inner vertex:
$Y_0=\{\|\}$. 
There is one tree with one inner vertex
$Y_1=\{\Y\}$, two trees with two inner vertices
$Y_2=\{\deuxun,\deuxdeux\}$,
five trees with three inner vertices
$Y_3=\{\troisun,\troisdeux,\troistrois,\troisquatre,\troiscinq\}$
and fourteen with four inner vertices
$Y_4=\{\quad\quatreun,\quad\quatredeux,\quad\quatretrois,\,\,\quatrequatre,\quatrecinq,
  \,\,\,\quatresix,\,\,\quatresept,\,\,\quatrehuit,\,\,\,\quatreneuf,\,\,\,\quatredix,
  \,\,\quatreonze, \,\,\quatredouze,\,\,\quatretreize,\,\,\quatrequatorze\}$.
The vertical line of a tree is called the \emph{root}.
The trees of $Y_n$ (with $n>0$) can be built from
smaller trees by the following relation
$Y_n=\{ t_1 \vee t_2 \,:\, t_1 \in Y_k, t_2 \in Y_{n-k-1},\, 
k=0,\dots,n-1\}$,
where $t_1 \vee t_2$ is the grafting of trees
$t_1$ and $t_2$, by which the roots of $t_1$ and $t_2$
are brought together and a new root is grown from their juncture.
Pictorially:
\begin{eqnarray*}
s \vee t &= \lrgraft{s}{t}.
\end{eqnarray*}
For example, $\|\vee\|=\Y$,
$\|\vee\Y=\deuxdeux$, $\Y \vee \Y= \troistrois$.
Except in figure~\ref{figarbre},
the vertices are not drawn explicitly for notational convenience.
The inner vertices of a tree are the vertices to which
three edges (or two edges and the root) are incident, 
its leaves are its vertices
to which a single edge is incident. A leaf is oriented either
to the left or to the right.
Each tree of $Y_n$ has  $n$ inner vertices and $n+1$ leaves.
The order $|t|$ of a tree $t$ is the number of its inner vertices.
If $C_n$ denotes the number of elements of $Y_n$, the recursive
definition of $Y_n$ implies that $C_0=1$ and
$C_n=\sum_{k=0}^{n-1} C_k C_{n-k-1}$, so that
$C_n$ is a Catalan number.
The recursive definition of planar binary trees make them
very easy to generate with a computer.

\subsection{Recursive relation between trees and RS terms}
In ref.~\onlinecite{BMP}, we showed that the wave operator can be written as the
sum
\begin{eqnarray}
\Omega &=& P + \sum_{n=1}^\infty \sum_{t\in Y_n} \Omega_t,
\label{Omegat}
\end{eqnarray}
where
\begin{eqnarray}
\Omega_t &=& \sum_{ij}|i\rangle \omega^{ij}_t \langle j|,
\label{defOmegat}
\end{eqnarray}
with
$|j\rangle$ an eigenstate of $H_0$ in the model space and 
$|i\rangle$ out of it (i.e. $P|j\rangle=|j\rangle$ and $P|i\rangle=0$).
The scalars $\omega^{ij}_t$ are defined by an exceedingly
simple recursive relation: for
any tree $t$ different from $\|$, there
are two trees $t_1$ and $t_2$ such that $t=t_1\vee t_2$. Then,
\begin{eqnarray}
\omega_t^{ij}  &=& -\sum\limits_{k,l} 
  \frac{\omega^{ik}_{t_1} \langle k |V|l\rangle \omega^{lj}_{t_2}}
   {e_j- e_i},
\label{recursive}
\end{eqnarray}
with the special cases 
$\omega^{ik}_{t_1}=-\delta_{i,k}$ if $t_1=\|$ and
$\omega^{lj}_{t_2}=+\delta_{l,j}$ if $t_2=\|$.
It is clear from the definition that $\Omega_t$
is built from a product of $|t|$ matrix elements of $V$.
Note that, when $t_1=\|$, both $i$ and $k$ in 
$\omega^{ik}_{t_1}$ correspond to eigenstates outside the model space,
while when $t_2=\|$, both $i$ and $k$ in 
$\omega^{ik}_{t_1}$ correspond to eigenstates in the model space.

We now prove that the expansion
$\chi=\sum_{t\not=|} \Omega_t$ 
is a solution
of eq.~(\ref{Lindgrenchi}). 
First notice that taking the commutator ${[}\Omega_t,H_0{]}$ amounts to multiply
$\omega_t^{ij}$ by $e_j-e_i$. 
Let $t=t_1 \vee t_2$. Then, 
${[}\Omega_t,H_0{]}$ is equal to $QVP$,
or $QV\Omega_{t_2}$ or $-\Omega_{t_1}VP$
or $-\Omega_{t_1} V \Omega_{t_2}$, respectively, according
to whether
$t_1=t_2=|$, or $t_1=|$ and $t_2\not=|$, or
$t_1\not=|$ and $t_2=|$, or $t_1\not=|$ and $t_2\not=|$,
respectively. The result follows by summing
over all trees $t_1$ and $t_2$.

This very simple proof illustrates the fact that
non trivial results can be easily obtained because
trees encapsulate the recursive structure
of the RS series.

A more explicit version of the recursive relation (\ref{recursive})
will sometimes be needed.
For $t=t_1\vee t_2$, we define equivalently $\Omega_t$ by
\begin{eqnarray}
\Omega_{t} &=& 
    \sum_{ij}
\frac{| i^Q \rangle \langle i^Q |V| j^P\rangle\langle j^P|}
   {e_j- e_i}\quad\mathrm{if}\,\, t_1=\| \,\,\mathrm{and}\,\, t_2=\|,
  \label{rec11}\\
\Omega_t  &=&
    \sum_{ij}
\frac{|i^Q\rangle \langle i^Q |V Q\, \Omega_{t_2}|j^P\rangle \langle j^P|}
   {e_j- e_i}\quad\mathrm{if}\,\, t_1=\| \,\,\mathrm{and}\,\,
t_2\not=\|,\label{rec1t}\\
\Omega_t  &=& -\sum_{ij}
  \frac{|i^Q\rangle\langle i^Q|\Omega_{t_1} P V|j^P\rangle\langle j^P|}
   {e_j- e_i}\quad\mathrm{if}\,\, t_1\not=\| \,\,\mathrm{and}\,\,
t_2=\|,\label{rect1}\\
\Omega_t  &=& -\sum_{ij}
  \frac{|i^Q\rangle\langle i^Q|\Omega_{t_1} P V Q\,\Omega_{t_2}|j^P\rangle
   \langle j^P|}
   {e_j- e_i}\quad\mathrm{if}\,\, t_1\not=\| \,\,\mathrm{and}\,\,
t_2\not=\|,\label{rectt}
\end{eqnarray}
where the exponent of $|j^P\rangle$ and $|i^Q\rangle$
means that state 
$|j\rangle$ is a basis state of $M$ and
$|i\rangle$ a basis state of the complementary of $M$.

\subsection{Examples}
As examples, we give all $\Omega_t$ for $|t|$=2 and 3,
the value of $\Omega_t$ for $|t|=1$ (i.e. $t=\Y$) being given by
eq.~(\ref{rec11})).
\begin{eqnarray*}
\Omega_t &=& - 
      \sum_{i_1,i_2,i_3} 
     \frac{|i_1^Q \rangle \langle i_1^Q|V|i_2^P\rangle
                   \langle i_2^P|V|i_3^P\rangle\langle i_3^P|}
                     {(e_{i_2}-e_{i_1})(e_{i_3}-e_{i_1})}
\quad\mathrm{for}\quad t=\deuxun,\\
\Omega_t &=&
      \sum_{i_1,i_2,i_3} 
     \frac{|i_1^Q \rangle \langle i_1^Q|V|i_2^Q\rangle
                   \langle i_2^Q|V|i_3^P\rangle\langle i_3^P|}
                     {(e_{i_3}-e_{i_2})(e_{i_3}-e_{i_1})}
\quad\mathrm{for}\quad t=\deuxdeux,\\
\Omega_t &=& 
      \sum_{i_1,i_2,i_3,i_4} 
     \frac{|i_1^Q \rangle \langle i_1^Q|V|i_2^P\rangle
                   \langle i_2^P|V|i_3^P\rangle
                   \langle i_3^P|V|i_4^P\rangle\langle i_4^P|}
             {(e_{i_2}-e_{i_1})(e_{i_3}-e_{i_1})(e_{i_4}-e_{i_1})}
\quad\mathrm{for}\quad t=\troisun,\\
\Omega_t &=& - 
      \sum_{i_1,i_2,i_3,i_4} 
     \frac{|i_1^Q \rangle \langle i_1^Q|V|i_2^Q\rangle
                   \langle i_2^Q|V|i_3^P\rangle
                   \langle i_3^P|V|i_4^P\rangle\langle i_4^P|}
             {(e_{i_3}-e_{i_1})(e_{i_3}-e_{i_2})(e_{i_4}-e_{i_1})}
\quad\mathrm{for}\quad t=\troisdeux,\\
\Omega_t &=& - 
      \sum_{i_1,i_2,i_3,i_4} 
     \frac{|i_1^Q \rangle \langle i_1^Q|V|i_2^P\rangle
                   \langle i_2^P|V|i_3^Q\rangle
                   \langle i_3^Q|V|i_4^P\rangle\langle i_4^P|}
             {(e_{i_2}-e_{i_1})(e_{i_4}-e_{i_3})(e_{i_4}-e_{i_1})}
\quad\mathrm{for}\quad t=\troistrois,\\
\Omega_t &=& - 
      \sum_{i_1,i_2,i_3,i_4} 
     \frac{|i_1^Q \rangle \langle i_1^Q|V|i_2^Q\rangle
                   \langle i_2^Q|V|i_3^P\rangle
                   \langle i_3^P|V|i_4^P\rangle\langle i_4^P|}
             {(e_{i_3}-e_{i_2})(e_{i_4}-e_{i_2})(e_{i_4}-e_{i_1})}
\quad\mathrm{for}\quad t=\troisquatre,\\
\Omega_t &=& 
      \sum_{i_1,i_2,i_3,i_4} 
     \frac{|i_1^Q \rangle \langle i_1^Q|V|i_2^Q\rangle
                   \langle i_2^Q|V|i_3^Q\rangle
                   \langle i_3^Q|V|i_4^P\rangle\langle i_4^P|}
             {(e_{i_4}-e_{i_3})(e_{i_4}-e_{i_2})(e_{i_4}-e_{i_1})}
\quad\mathrm{for}\quad t=\troiscinq.
\end{eqnarray*}
By looking at these examples,
one may wonder whether some denominators could
take the value zero.
We show now by induction that this never happens. In other words,
$\Omega_t$ is never singular.
This is true for $t=\Y$ because, in eq.~(\ref{rec11}),
$|i\rangle$ belongs to $M$ and 
$|j\rangle$ does not belong to $M$. Thus,
$|e_i-e_j|$ is greater than the gap between the 
model space and the rest of the spectrum.
Assume now that no $\Omega_t$ is singular
for trees with $|t|\le n$. Take a tree
$t$ with $|t|=n+1$. Then, $t=t_1\vee t_2$,
with $|t_1|\le n$ and $|t_2|\le n$.
Assume that neither $t_1$ nor $t_2$ is equal to the
root $\|$. Then, eq.~(\ref{rectt}) and the previous
remark about $|e_i-e_j|$ shows that $\Omega_t$
is not singular. The same is true if 
$t_1$ or $t_2$ is equal to the root.
Thus, the denominator $\Omega_t$ can never
take the value 0. Again, the proof is made very simple
by the recursive structure of $\Omega_t$.

\subsection{Direct relation between trees and RS terms}
\begin{figure}
\includegraphics[width=8.0cm]{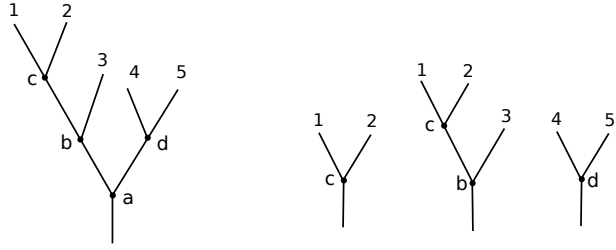}
\caption{A tree with its labelled leaves and its three proper subtrees.
The inner vertices are labelled $a$, $b$, $c$ and $d$.
The couples $(l(v),r(v))$ are $(1,5)$, $(1,3)$, $(1,2)$ and $(4,5)$
for $v=a,b,c$ and $d$, respectively.
\label{figarbre}}
\end{figure}

The recursive relation~(\ref{recursive}) is very useful
to derive proofs, but we also need a non-recursive
expression for the terms of the RS series. 
We now give an explicit relation between $t$
and $\Omega_t$~\cite{BMP}.
This construction can be followed in figure~\ref{figarbre} for
$t=\quatresix$.
Consider a tree $t$ with $|t|=n$ and number its leaves from
1 for the leftmost leaf to $n+1$ for the rightmost one.
The numerators in the expansion of $\Omega_t$ are
$|i_1\rangle\langle i_1|V|i_2\rangle\dots
\langle i_n|V|i_{n+1}\rangle\langle i_{n+1}|$.
The state $|i_k\rangle$ belongs to $M$ 
(i.e. is $|i_k^P\rangle$) if leaf $k$
is oriented to the right and does not belong to $M$ 
(i.e. is $|i_k^Q\rangle$) if leaf $k$
is oriented to the left.
In the example of fig.~\ref{figarbre}, the denominator is
$|i_1^Q\rangle \langle i_1^Q |V|i_2^P\rangle
        \langle i_2^P |V|i_3^P\rangle
        \langle i_3^P |V|i_4^Q\rangle
        \langle i_4^Q |V|i_5^P\rangle\langle i_5^P|$

Then, for each inner vertex $v$ of $t$, take the subtree
$t_v$ for which $v$ is just above the root. In other words,
$t_v$ is obtained by chopping the edge below $v$ and
considering the edge dangling from $v$ as the root of $t_v$.
In the example of fig.~\ref{figarbre}, we have four inner vertices labelled
$a$, $b$, $c$, $d$. For vertex $a$, the subtree $t_a$ is
the full tree $t$. For the other inner vertices the subtrees
are proper (i.e. different from $t$) and are given in
fig.~\ref{figarbre}.
For each tree $t_v$, denote by $l(v)$ the index of its leftmost
leaf and by $r(v)$ that of its rightmost leaf. 
Then, the denominator is the product of
$e_{i_{r(v)}}-e_{i_{l(v)}}$, where $v$ runs over the $n$
inner vertices  of the tree.
In the example of fig.~\ref{figarbre}, $t_a$ gives the denominator 
$(e_{i_5}-e_{i_1})$, $t_b$ gives $(e_{i_3}-e_{i_1})$,
$t_c$ gives $(e_{i_2}-e_{i_1})$ and $t_d$ gives $(e_{i_5}-e_{i_4})$.
Finally, the whole fraction is summed over
$i_1\dots i_{n+1}$ and multiplied by $(-1)^{d-1}$,
where $d$ is the number of leaves pointing to the right.
It can be easily proved that the above construction satisfies the recursive
equation (\ref{recursive}).
If we take the example of the tree $t$ of fig.~\ref{figarbre} and consider for example the inner vertex labelled by $b$, we get a contribution $(e_{i_3}-e_{i_1})$ to the denominator. The total term is
\begin{eqnarray*}
\Omega_t & = & (-1)^2
  \sum_{i_1 i_2 i_3 i_4 i_5}
  \frac{|i_1^Q\rangle \langle i_1^Q |V|i_2^P\rangle
        \langle i_2^P |V|i_3^P\rangle
        \langle i_3^P |V|i_4^Q\rangle
        \langle i_4^Q |V|i_5^P\rangle\langle i_5^P|}
      { (e_{i_5} - e_{i_1})
     (e_{i_2} - e_{i_1})
     (e_{i_3} - e_{i_1})
     (e_{i_5} - e_{i_4})}.
\end{eqnarray*}

As far as we know, this tree representation provides the first
description of the general term of the RS series for
quasidegenerate systems.

\section{Resummations}
The tree representation is useful, not only to prove
properties of the RS series, but also to derive new resummations
of it. Indeed, any way to write the set of trees as a composition
of subtrees gives rise to a resummation of the
RS series.

\subsection{Summation over left combs}\label{sub1}

For any tree $t$, we define the sequence
$t_n$ by $t_0=t$, $t_{n+1}=t_n\vee |$.
For example, if $t=\deuxdeux$, we have
$t_1=\troisdeux$, $t_2=\quatredeux$.
For any integer $n$, $t_n$ is obtained by grafting
$t$ on the leftmost leaf of a tree that has
a single leaf oriented to the left and $n$ leaves oriented to the
right (such a tree is called a left comb whereas $t_n$, $n>0$ is called 
a left comb grafting. Notice the uniqueness rewriting property: an arbitray 
tree $u$ can be rewritten uniquely as $t_n$, where $t=|$ or $t\not= v\vee |$.
The sum over graftings on left combs is made by defining
\begin{eqnarray*}
\Omega'_t &=& \Omega_t+\sum_{n=1}^\infty \Omega_{t_n}.
\end{eqnarray*}
We can now calculate $\Omega'_t$ in terms of $\Omega_t$.
We first define 
\begin{eqnarray*}
G^0_P(z) &=& P(H_0-z)^{-1}P = \sum_j \frac{|j^P\rangle\langle
j^P|}{e_j-z}.
\end{eqnarray*}
Then, we use Kvasni{\v{c}}ka's trick and denote by
$Q_i= |i^Q\rangle\langle i^Q|$ the projector onto the
eigenspace of $H_0$ with eigenvalue $e_i$ outside the model space,
so that $Q=\sum_i Q_i$ and eq.~(\ref{rect1}) can be rewritten
$\Omega_{t_1} = -\sum_i Q_i \Omega_t PV G^0_P(e_i)$.
By repeating this argument we obtain
\begin{eqnarray*}
\Omega'_t &=& \sum_i Q_i \Omega_t
\Big(P - PV G^0_P(e_i) + PV G^0_P(e_i)PV G^0_P(e_i)-\dots
\\&=&
 \sum_i Q_i \Omega_t
\Big(P + PV G^0_P(e_i)\Big)^{-1}.
\end{eqnarray*}
The map $P+PV G^0_P(e_i)$ goes from the model space to itself and
the inverse is taken within the model space.
The dimension of the model space is generally small and
the numerical calculation of the inverse is fast.
We transform
\begin{eqnarray*}
\Big(P + PV G^0_P(e_i)\Big)^{-1}
&=& \Big((H_0 P -e_i P + PV P)G^0_P(e_i)\Big)^{-1}
=  ( H_0 P-e_i P) (PH P -e_i P)^{-1},
\end{eqnarray*}
where $PHP=P(H_0+V)P$ and the inverse is again from $M$ to $M$.
To conclude, we assume that 
$t=u\vee v$, where $u$ and $v$ are different from $\|$.
Then,
\begin{eqnarray*}
\Omega'_t &=& \sum_i Q_i  \Omega_u
V \Omega_v G^0_P(e_i) (P H_0-e_i P) (PH P -e_i P)^{-1}
\\&=&
 \sum_i Q_i \Omega_u
V \Omega_v (PH P -e_i P)^{-1}
=
 \sum_{ij} \frac{|i\rangle\langle i| \Omega_u
V \Omega_v \brap{j}\ketp{j}}{\enep{j} -e_i},
\end{eqnarray*}
where $\brap{j}$ and $\enep{j}$ are the eigenstates
and eigenvalues of $PHP$, so that
$P=\sum_j \brap{j}\ketp{j}$.
In other words, summing over all left combs amounts to 
replacing the eigenstates of $PH_0P$ by
the eigenstates of $PHP$.

\subsection{Accelerarated summation over left combs}
The formula we obtained for $\Omega'_{t}$ suggests to look for 
another expansion of the RS series involving only the eigenstates 
and eigenvalues of $P(H_0+V)P$. 
The uniqueness rewriting property implies that
 $\Omega=P + \Omega'_{\Y}+
\sum_t \Omega'_t$, 
where the sum is over all trees $t$  with two or more inner vertices and such that $t\not= u\vee |$, but this resummation can still be improved.

We rewrite the Kvasni{\v{c}}ka-Lindgren equation.
\begin{eqnarray*}
{[}\chi,H_0{]} &=& 
\sum_i Q_i \chi H_0 - H_0 \sum_i Q_i \chi
=
\sum_i Q_i \chi (H_0 - e_i).
\end{eqnarray*}
Thus, eq.~(\ref{Lindgrenchi}) becomes
$\sum_i Q_i \chi (H_0 - e_i) = QVP +QV\chi -\chi V P - \chi V \chi$
that is,
\begin{eqnarray}
\sum_i Q_i \chi P(H_0 - e_i+V)P &=& QVP +QV\chi  - \chi V \chi,
\label{Lindgrenchip}
\end{eqnarray}
or, for all the eigenvalues $e_i$ of $H_0$ outside the model space:
\begin{eqnarray*}
Q_i\chi P=Q_iVG_P(e_i)+Q_iV\chi G_P(e_i) - Q_i\chi V \chi G_P(e_i),
\end{eqnarray*}
where
\begin{eqnarray*}
G_P(z) &=& (PH_0P + PVP -zP)^{-1} = \sum_j \frac{\brap{j}\ketp{j}}{\enep{j}-z}.
\end{eqnarray*}

This equation is similar to the Kvasni{\v{c}}ka-Lindgren equation and 
can be solved graphically and recursively by the same process.
Notice that the leading term of the recursive expansion of $\chi$ is now 
$\Omega_{\Y}'=\sum\limits_iQ_iVG_P(e_i)$. 

Let us call a tree $t$ right-normalized if there is no edge in $t$ 
such that the tree $t'$ obtained by pruning the tree $t$ at that edge 
can be written $t_1\vee \|$ with $t_1\not= \|$. The only left comb
possibly contained in a right-normalized tree is $\Y$.
The number of right-normalized trees is 
considerably smaller than the number of trees.
For $n$=1, 2, 3 and 4 the number of right-normalized trees
is 1 ($\Y)$, 1 ($\deuxdeux$), 2 ($\troistrois$, $\troiscinq$)
and 4 ($\,\,\quatresept$, $\,\,\,\quatreneuf$, $\,\,\quatredouze$,
$\,\quatrequatorze$)
instead of 1, 2, 5, 14.
The right-normalized trees are enumerated by the
Motzkin numbers~\cite{Donaghey-77}.

The solution of eq.~(\ref{Lindgrenchip} ) is then
$\Omega = P + \sum_t \hatOmega_t$, where $t$ runs over right-normalized 
trees and $\hatOmega_t$ can be defined recursively by 
\begin{eqnarray}
\hatOmega_{t} &=& \sum\limits_{i \notin M,j\in M}
\frac{|i\rangle\langle i |V\brap{j}\ketp{j}}
   {\enep{j}- e_i}\quad\mathrm{if}\,\, t_1=\| \,\,\mathrm{and}\,\, t_2=\|,
  \label{recp112}\\
\hatOmega_t  &=& \sum\limits_{i \notin M,j\in M}
\frac{|i\rangle\langle i |V\hatOmega_{t_2}\brap{j}\ketp{j}}
   {\enep{j}- e_i}\quad\mathrm{if}\,\, t_1=\| \,\,\mathrm{and}\,\,
t_2\not=\|,\label{recp1t2}\\
\hatOmega_t  &=& -\sum\limits_{i \notin M,j\in M}
  \frac{|i\rangle\langle i |\hatOmega_{t_1} V \hatOmega_{t_2}\brap{j}\ketp{j}}
   {\enep{j}- e_i}\quad\mathrm{if}\,\, t_1\not=\| \,\,\mathrm{and}\,\,
t_2\not=\|.\label{recptt2}
\end{eqnarray}

Once again, this approach reduces the number. It can be expected
that this resummation considerably accelerates the convergence of the series.
Indeed, when the model space is well chosen, the matrix elements
of $QVP$ are smaller than those of $PVP$. In this expression,
all terms involving powers of $PVP$ have been resummed.

\subsection{Alternative accelerated tree expansion}\label{secondsect}
A similar resummation, that we omit, can be obtained by summing over
right combs. However, it is not so practical as the
previous ones because it is usually not easy to invert
$QHQ-Qz$. If this inversion is possible, then we can 
simultaneously sum over right and left combs, at least if we 
assume that a finite gap exists between the eigenvalues of $QHQ$ and $PHP$.
We make this assumption, proceed as before and transform the 
Kvasni{\v{c}}ka-Lindgren equation.
We define
$\Pp{j}=\brap{j}\ketp{j}$ the 
projector onto the eigenspace of $PHP$ with energy
$\enep{j}$ and use similar notations for the projectors and 
energy levels of $QHQ$. Thus, $P=\sum_j \Pp{j}$ and
we can rewrite the lhs of eq.~(\ref{Lindgrenchip}) as
\begin{eqnarray*}
\sum_i Q_i \chi P(H_0 - e_i+V)P &=& 
\sum_{ij} Q_i \chi P(H_0 - e_i+V) \Pp{j}
=\sum_{ij} Q_i \chi (\enep{j} - e_i) \Pp{j}
=\sum_{j} Q (\enep{j}-H_0) \chi \Pp{j}.
\end{eqnarray*}
Thus eq.~(\ref{Lindgrenchip}) becomes
$\sum_{j} Q (\enep{j}-H_0) \chi \Pp{j} 
=
QVP +QV\chi  - \chi V \chi$, or
\begin{eqnarray*}
\sum_{j} Q (\enep{j}-H_0- V) \chi \Pp{j} 
&=&
QVP  - \chi V \chi.
\end{eqnarray*}
For all eigenvalues $\overline e_j$ of $PHP$, this gives us
$$\chi \overline P_j=S(\overline e_j)V\overline P_j-S(\overline e_j)\chi 
  V\chi \overline P_j,$$
where we set: 
$$S(z)=(z Q - QHQ)^{-1}.$$

We  can write
directly the wave operator as a sum parametrized by trees:
$\Omega = P + \sum_{t\ge |} \tilde\Omega_t$, where 
$\tilde\Omega_t$ is defined recursively for all trees by 
\begin{eqnarray}
\tilde\Omega_{t} &=& \sum\limits_{i \notin M,j\in M}
\frac{\brap{i}\ketp{i}V\brap{j}\ketp{j}}
   {\enep{j}- \enep{i}}
= \sum_{j\in M} S(\enep{j})QV \Pp{j}
\quad\mathrm{if}\,\, t=\|,
  \label{recq11}\\
\tilde\Omega_t  &=& -\sum\limits_{i \notin M,j\in M}
  \frac{\brap{i}\ketp{i}\tilde\Omega_{t_1} V \tilde\Omega_{t_2}\brap{j}\ketp{j}}
   {\enep{j}- \enep{i}}
= -\sum_{j\in M} S(\enep{j})
  \tilde\Omega_{t_1} V \tilde\Omega_{t_2}\Pp{j}
\quad\mathrm{if}\,\, t=t_1\vee t_2.\label{recqtt}
\end{eqnarray}

As for the one involving a direct resummation over right combs,
this resummation is not as practical as the
previous one because the eigenvalues of
$QHQ$ are generally not known.
However, this algorithm provides a way to express the
eigenvectors of a matrix
$M = \left(\begin{array}{cc} 
    A & C  \\
    C^\dagger & B 
  \end{array}\right)$ 
in terms of the eigenvalues and eigenvectors of $A$ and $B$.

\subsection{Fourth resummation}
We use now another representation of trees in terms of trees. 
Since any tree $t$ decomposes uniquely as $t=t_1\vee t_2$, 
a recursion on the left hand side of the decomposition shows that
there is a unique integer $k$
and $k$ unique trees $ v_i$ (with $i=1,\dots,k$) such that
\begin{eqnarray*}
t = \calL_k(v_1,\dots,v_k),
\end{eqnarray*}
where
\begin{eqnarray*}
\calL_k(v_1,\dots,v_k) &=&
\Big(\big(\dots (\|\vee v_1)\vee\dots\big)\vee v_{k-1}\Big)\vee v_k.
\end{eqnarray*}

In words, 
$\calL_k(v_1,\dots,v_k)$ is obtained by taking a left comb with $k$ 
right leaves and grafting $v_k$ on the lowest right leaf,
$v_{k-1}$ on the next one, up to $v_1$ on the highest right leaf.

If we sum over $k$ and over all trees $v_i$ we find
\begin{eqnarray}
F &=& \| + \sum_{k=1}^\infty \calL_k(F,\dots,F),
\label{eqF}
\end{eqnarray}
where $F$ stands for the formal sum of all trees.

We can plug the expansion of the wave operator
of section~\ref{secondsect} in this equation and get
for $t=\calL_k(v_1,\dots,v_k)$:
\begin{eqnarray*}
\tilde\Omega_t &=& (-1)^k \sum_{j_1\dots j_{k+1}}
  S(\enep{j_{k+1}})\dots S(\enep{j_1})
   QV\Pp{j_1} V \tilde\Omega_{v_1}\Pp{j_2} V \dots V \tilde\Omega_{v_k}\Pp{j_{k+1}}.
\end{eqnarray*}
Therefore, eq.~(\ref{eqF}) gives us the non-perturbative equation
\begin{eqnarray}
\chi &=& \sum_j S(\enep{j})V\Pp{j}  + \sum_{k=1}^\infty (-1)^k \sum_{j_1\dots j_{k+1}}
  S(\enep{j_{k+1}})\dots S(\enep{j_1})
   QV\Pp{j_1} V \chi \Pp{j_2} V \dots V \chi \Pp{j_{k+1}}.
\label{chiLk}
\end{eqnarray}

If we use instead the first expansion Eq~(\ref{Omegat}) of the wave operator 
and plug the corresponding recursive formulas in $\calL_k(v_1,\dots,v_k)$, 
we get
$$\Omega_t=(-1)^{k-1}\sum_{j_1\dots j_{k+1}}S_0(e_{j_{k+1}})...S_0(e_{j_{1}})
QVP_{j_1}V\Omega_{v_1}P_{j_2}V\Omega_{v_2}...V\Omega_{v_k}P_{j_{k+1}}$$
where $\Omega_|:=P$ and $S_0(z):=(zQ-QH_0Q)^{-1}$, so that:
$$\Omega =P+\sum\limits_{k=1}^\infty (-1)^{k-1}
\sum\limits_i(...((Q_iVG_p^0(e_i))V\Omega G_p^0(e_i))...V\Omega G_p^0(e_i)).$$

We believe that these relations are new.
Similar expansions would follow by considering the symmetric expansion 
of trees (using the grafting operation of a tree on the right-most leaf 
of another tree instead of the grafting on the left-most tree).

\subsection{Relation with previous works}
The previous works on quasidegenerate perturbation theory
correspond to a summation which is symmetric to our first alternative expansion (in the sense that they focus on the operators $(e_i-QHQ)$).
By following a line suggested by 
Kvasni{\v{c}}ka and Lindgren~\cite{Kvasnicka-74,Lindgren74},
several authors transformed the Kvasni{\v{c}}ka-Lindgren equation into
\begin{eqnarray*}
\sum_i (e_i -QHQ) \chi P_i &=&
  QVP - \chi V P - \chi V \chi.
\end{eqnarray*}
A resummation (similar, but slightly more involved than the one leading to eq.~(\ref{chiLk})) gives the
equation~\cite{Hose-79,SO2}
\begin{eqnarray*}
\chi &=& \sum_j S(e_j)V P_j  + 
   \sum_{k=1}^\infty (-1)^k \sum_{j_1\dots j_{k+1}}
  S(e_{j_{k+1}})\dots S(e_{j_1})
   V P_{j_1} (V+V\chi)P_{j_2} (V+V\chi) \dots (V+V \chi) P_{j_{k+1}},
\end{eqnarray*}
which is a generalization of the
degenerate case~\cite{Cloizeaux-60,Brandow}.
Suzuki and Okamoto further solved this for $\chi$ but their
rather complex result was not applied to concrete problems, as far
as we know.
Up to a left/right symmetry, their result is similar to our first alternative expansion, excepted for the fact
that $PVP$ and $PV S(e)VP$ are grouped in a single term.

The main practical difference between the result obtained by
Suzuki and Okamoto and our second alternative expansion
is the fact that we do not consider the term $PVP$ as
a perturbation, we treat it exactly. This is important
because, when the model space is well chosen, $PVP$
is larger than the non-diagonal terms $PVQ$.

\section{Green function of degenerate systems}
In this section, we discuss a question related to
the Green function of degenerate systems.
Consider a Hamiltonian $H_0$ with a degenerate energy $e_0$.
The eigenstates of $H_0$ with energy $e_0$ span
a vector space $M$. The projector onto the model space $M$
is denoted by $P$. 
In a series of recent papers~\cite{BSP,BPS-PRL,BPS}, we proved by
non-perturbative methods that there are eigenstates $|i\rangle$ of $H_0$,
called the \emph{parent states}, such that
the usual Gell-Mann and Low wavefunction has a
well-defined limit when the adiabatic parameter $\epsilon$
goes to zero:
\begin{eqnarray*}
|\Psi_{\mathrm{GML}}\rangle &=& 
  \lim_{\epsilon\to0} \frac{U_\epsilon(0,-\infty)|i\rangle}
{\langle i | U_\epsilon(0,-\infty)|i\rangle},
\end{eqnarray*}
where $U_\epsilon(t,t')$ is the evolution operator in the
interaction picture.
Morover, we showed that the parent states
$|i\rangle$ are eigenstates of $H_0$ (with energy $e_0$)
and are also eigenstates of $PVP$. 
As a consequence, $\langle i|V|j\rangle=0$ for $i\not= j$
if $|i\rangle$ and $|j\rangle$ are parent states.

Since the parent states solve the problem in a non-perturbative
approach, one might be tempted to use them in the perturbative one.
In other words, we pick up a parent state, say $|0\rangle$, 
and we calculate the Rayleigh-Schr\"odinger series corresponding
to  the projector $|0\rangle\langle 0|$.
However, as noticed by T{\'o}th~\cite{Toth-10}, a problem
appears in the perturbative expansion.
This problem can be illustrated by a simple example:
for $t=\deuxun$, we have~\cite{BMP}
\begin{eqnarray*}
\Omega_t &=& -
      \sum_{i\not=0} 
     \frac{|i\rangle \langle i|V|0\rangle
                   \langle 0|V|0\rangle\langle 0|}
          {(e_{0}-e_{i}+i\epsilon)(e_{0}-e_{i}+2i\epsilon)},
\end{eqnarray*}
where $\epsilon$ is the adiabatic switching operator,
which tends to zero at the end of the calculation.
Now, if $|i\rangle$ belongs to the model space,
then $\langle i|V|0\rangle=0$ (because $i\not=0$)
and the expression converges although
$e_i=e_0$ could have brought a problem.
In other words, using a basis of parent
states for $M$ has made the expression convergent.
However, this trick does not always work.
Indeed, for $t=\deuxdeux$, we have~\cite{BMP}
\begin{eqnarray*}
\Omega_t &=& 
      \sum_{i\not=0,j\not=0} 
     \frac{|i\rangle \langle i|V|j\rangle
                   \langle j|V|0\rangle\langle 0|}
          {(e_{0}-e_{j}+i\epsilon)(e_{0}-e_{i}+2i\epsilon)}.
\end{eqnarray*}
If $|i\rangle$ belongs to the model space and $|j\rangle$ is
out of it, then we have $e_i=e_0$ and the limit $\epsilon\to0$
is not defined because nothing insures that
$\langle i|V|j\rangle=0$ or $\langle j|V|0\rangle=0$.
In other words, the convergence problem was solved at
the non-perturbative level but remains at the perturbative one,
so that the perturbative expansion has to be resummed in
a proper way.

Now we show how to solve this problem by using a trick related to 
the Hamiltonian shift proposed by Silverstone~\cite{Silverstone-71-JCP}.
We assume that
$e'_i=e_i + \langle i|V|i\rangle$ are nondegenerate.
Then, we rewrite
$H=H_0+V=H'_0+V'$, where
\begin{eqnarray*}
H'_0 &=& H_0+\sum_{i\in M} |i\rangle\langle i|V|i\rangle\langle i|,\\
V' &=& V - \sum_{i\in M} |i\rangle\langle i|V|i\rangle \langle i|.
\end{eqnarray*}
We build the RS series for $H'_0$ and $V'$ with the one-dimensional
model space $M$ spanned by $|0\rangle$.
Thus, $P'=|0\rangle\langle 0|$.  This gives us $P' V' P'=0$.
As a consequence, $\Omega_t=0$
if $t=t_1\vee\|$ with $t_1\not=\|$.
We can write $Q'=Q_0+Q$, where $Q$
is the projector corresponding to the initial problem and
$Q_0=P-P'$ is the projector onto the basis states
of $M$ different from $|0\rangle$.
Then, we have $Q_0 V' P'=0$ and $Q V'=QV$.
This gives us  $Q'V'P'=QV'P'=QVP'$,
which simplifies the evaluation of $\Omega_t$
for $t=\Y$.
Similarly, $P'V'Q'=P'VQ$ simplifies the evaluation
of $\Omega_t$ for $t=t_1\vee t_2$.
Finally, the identity
$Q_0 V' Q_0=0$ gives us
$Q'V'Q'=Q_0 V Q + Q V Q_0 + Q V Q$
for the evaluation of $\Omega_t$ with
$t=\|\vee t_2$.

This gives us the following recursive expression
for $t=t_1\vee t_2$.
\begin{eqnarray*}
\Omega_{t} &=& 
    \sum_{i}
\frac{| i^Q \rangle \langle i^Q |V| 0\rangle\langle 0|}
   {e'_0- e_i}\quad\mathrm{if}\,\, t_1=\| \,\,\mathrm{and}\,\, t_2=\|,
  \\
\Omega_t  &=&
    \sum_{i}
\frac{|i^Q\rangle \langle i^Q |V (Q+Q_0)\, \Omega_{t_2}|0\rangle \langle 0|}
   {e'_0- e_i}
+
    \sum_{i}
\frac{|i^{Q_0}\rangle \langle i^{Q_0} |V Q\, \Omega_{t_2}|0\rangle \langle 0|}
   {e'_0- e'_i}
\quad\mathrm{if}\,\, t_1=\| \,\,\mathrm{and}\,\,
t_2\not=\|,\\
\Omega_t  &=& 0 \quad\mathrm{if}\,\, t_1\not=\| \,\,\mathrm{and}\,\,
t_2=\|,\\
\Omega_t  &=& -\sum_{i}
  \frac{|i^Q\rangle\langle i^Q|\Omega_{t_1}|0\rangle
    \langle 0| V Q\,\Omega_{t_2}|0\rangle \langle 0|}
   {e'_0- e_i}
 -\sum_{i}
  \frac{|i^{Q_0}\rangle\langle i^{Q_0}|\Omega_{t_1}|0\rangle
    \langle 0| V Q\,\Omega_{t_2}|0\rangle \langle 0|}
   {e'_0- e'_i}\quad\mathrm{if}\,\, t_1\not=\| \,\,\mathrm{and}\,\,
t_2\not=\|.
\end{eqnarray*}
The recursive expression shows that all terms are
well defined if all 
$e'_i=e_i + \langle i|V|i\rangle$ are different
and if $e'_0$ is different from the energies $e_j^Q$
out of the model space.

\section{Continued fractions}
We discuss here some continued-fraction resummation of
the RS series. Such a (generalized) continued fraction formula was found 
to be very efficient for calculating nuclear properties~\cite{Coraggio}.
The combinatorial structure of continued fractions was studied in
detail by Flajolet~\cite{Flajolet-80,Flajolet-06}. 
On the other hand, Lee and Suzuki derived a continued fraction
expression for $\chi$ and the effective Hamiltonian
for a degenerate system~\cite{LeeSuzuki,Suzuki-80}.
Other implementations of continued fractions for pertubation theory
can be found in the 
literature~\cite{Young-57,Feenberg-58,Scofield-72,Makowksi-85,Swain-94}.

\subsection{The Suzuki-Lee formula}
Suzuki and Lee~\cite{Suzuki-80}
 start from the Kvasnicka-Lindgren equation
\begin{eqnarray*}
{[}\chi,H_0{]} &=& QVP + QV\chi -\chi V P -\chi V \chi.
\end{eqnarray*}
For a degenerate system $P H_0=e_0 P$, so that $\chi H_0= e_0 \chi$ and
$(e_0-H_0) \chi = QVP + QV\chi -\chi V P -\chi V \chi$.
This equation is then reordered into
$(e_0-Q H Q+ \chi VQ) \chi = QVP  -\chi  V P$.
They consider the iterative equation (see \cite{Suzuki-80}, eq.~(3.27) p.~2102).
\begin{eqnarray*}
(e_0-Q H Q+ \chi_{n-1} VQ) \chi_n &=& QVP  -\chi_{n-1}  V P,
\end{eqnarray*}
with the boundary condition $\chi_0=0$.
In other words
\begin{eqnarray*}
\chi_n &=& 
(e_0-Q H Q+ \chi_{n-1} PVQ)^{-1} (QVP  -\chi_{n-1}  V P).
\end{eqnarray*}

\subsection{A new continued fraction expansion}
The Suzuki-Lee formula has two drawbacks: it is restricted to
degenerate systems~\cite{suzuki-quasi-degenerate}
 and it requires the inversion of $e_0-Q H Q+ \chi_{n-1} PVQ$,
which is usually infinite dimensional.
To solve these two problems, we start from
eq.~(\ref{Lindgrenchip}) and, by using $Q_i Q=Q_i$, we rewrite it
$Q_i \chi P(H - e_i)P = Q_iVP +Q_iV\chi  - Q_i\chi V \chi$.
We transform this equation into
\begin{eqnarray}
Q_i \chi \Big( P(H - e_i)P + P V \chi\Big) &=& Q_iVP +Q_iV\chi.
\label{Qichi}
\end{eqnarray}
Thus, we define the system of recursive equations
\begin{eqnarray*}
Q_i \chi_n &=& (Q_iVP +Q_iV\chi_{n-1})( PHP  - e_iP + P V \chi_{n-1})^{-1},\\
\chi_n &=& \sum_i Q_i \chi_n,
\end{eqnarray*}
with the boundary condition $\chi_0=0$.
This generalized continued fraction has convergence properties similar
to that of Lee and Suzuki, it is well-defined for quasi-degenerate systems
and the inverse is computationally easier because it is done within the model
space.

\section{Bijections}
\label{bijection-sect}
Many other combinatorial objects have been used to 
represent the terms of the RS series in non-degenerate
or degenerate cases.
Each of these representations is useful for specific applications.
It is therefore important to describe the relation between
the most important of them
(Bloch sequences, Dyck paths, braketings and non-crossing partitions)
and the trees.
Most of these representations are valid for degenerate systems. Thus,
we start by presenting the first terms of the RS series of degenerate systems,
where $e_0$ is the energy of the states of the model space.
We also give an operator representation of these terms, 
with $R=Q (e_0-H_0)^{-1} Q$.
\begin{eqnarray*}
\Omega_t &=& \sum_{i_1,i_2}\frac{|i_1^Q\rangle\langle i_1^Q|V|i_2^P\rangle
                             \langle i_2^P|}
                     {e_{0}-e_{i_1}}
  = RVP
\quad\mathrm{for}\quad t=\Y,\\
\Omega_t &=& - \sum_{i_1,i_2,i_3} \frac{|i_1^Q\rangle
                     \langle i_1^Q|V|i_2^P\rangle
                   \langle i_2^P|V|i_3^P\rangle\langle i_3^P|}
                     {(e_{0}-e_{i_1})(e_{0}-e_{i_1})}
   = - R^2 VPVP
\quad\mathrm{for}\quad t=\deuxun,\\
\Omega_t &=& \sum_{i_1 i_2 i_3}\frac{|i_1^Q\rangle\langle i_1^Q|V|i_2^Q\rangle
                   \langle i_2^Q|V|i_3^P\rangle\langle i_3^P|}
                     {(e_{0}-e_{i_2})(e_{0}-e_{i_1})}
  = RVRVP
\quad\mathrm{for}\quad t=\deuxdeux,\\
\Omega_t &=& \sum_{i_1 i_2 i_3 i_4}\frac{|i_1^Q\rangle
      \langle i_1^Q|V|i_2^P\rangle
                   \langle i_2^P|V|i_3^P\rangle
                   \langle i_3^P|V|i_4^P\rangle
                   \langle i_4^P|}
             {(e_{0}-e_{i_1})(e_{0}-e_{i_1})(e_{0}-e_{i_1})}
   = R^3 V P VP VP
\quad\mathrm{for}\quad t=\troisun,\\
\Omega_t &=&-\sum_{i_1 i_2 i_3 i_4}\frac{|i_1^Q\rangle
        \langle i_1^Q|V|i_2^Q\rangle
        \langle i_2^Q|V|i_3^P\rangle
        \langle i_3^P|V|i_4^P\rangle
        \langle i_4^P|
                   }
             {(e_{0}-e_{i_1})(e_{0}-e_{i_2})(e_{0}-e_{i_1})}
  = - R^2 V R V P V P
\quad\mathrm{for}\quad t=\troisdeux,\\
\Omega_t &=&-\sum_{i_1 i_2 i_3 i_4}\frac{|i_1^Q\rangle
        \langle i_1^Q|V|i_2^P\rangle
        \langle i_2^P|V|i_3^Q\rangle
        \langle i_3^Q|V|i_4^P\rangle
        \langle i_4^P|
                   }
             {(e_{0}-e_{i_1})(e_{0}-e_{i_1})(e_{0}-e_{i_3})}
  = - R^2 V P V R V P
\quad\mathrm{for}\quad t=\troistrois,\\
\Omega_t &=&-\sum_{i_1 i_2 i_3 i_4 }\frac{|i_1^Q\rangle
        \langle i_1^Q|V|i_2^Q\rangle
        \langle i_2^Q|V|i_3^P\rangle
        \langle i_3^P|V|i_4^P\rangle
        \langle i_4^P|
                   }
             {(e_{0}-e_{i_1})(e_{0}-e_{i_2})(e_{0}-e_{i_2})}
  = - R V R^2 V P V P
\quad\mathrm{for}\quad t=\troisquatre,\\
\Omega_t &=&\sum_{i_1 i_2 i_3 i_4}\frac{|i_1^Q\rangle
        \langle i_1^Q|V|i_2^Q\rangle
        \langle i_2^Q|V|i_3^Q\rangle
        \langle i_3^Q|V|i_4^P\rangle
        \langle i_4^P|
                   }
             {(e_{0}-e_{i_3})(e_{0}-e_{i_2})(e_{0}-e_{i_1})}
  = R V R V R V P
\quad\mathrm{for}\quad t=\troiscinq.
\end{eqnarray*}

\begin{table}
\centering
\begin{tabular}{|c||c|c|c|c|c|c|}\hline
$n$ & Tree & Bloch & Dyck & Bracketing & Partition & \cite{Olszewski-04}\\
\hline\hline
1 & \lY & (1) & \DY & ${*} o \rangle $ & $|1|$ & 1 \\
\hline\hline
2 & \ldeuxun & (20) & \Ddeuxun & $-{*} \langle o\rangle {*} o \rangle $ & $|12|$ & 2 \\
\hline
2 & \ldeuxdeux & (11) & \Ddeuxdeux & ${*}  o {*} o \rangle $
   & $|1|2|$ & 1 \\
\hline\hline
3 & \ltroisun & (300) & \Dtroisun 
   & ${*}  \langle o  \rangle {*} \langle o\rangle {*} o \rangle$
   & $|123|$ & 5 \\
\hline
3 & \ltroisdeux & (210) & \Dtroisdeux
   & $-{*}  \langle o  \rangle {*}  o {*} o \rangle$
   & $|13|2|$ & 2 \\
\hline
3 & \ltroistrois & (201) & \Dtroistrois
   & $-{*}  \langle o   {*}  o\rangle {*} o \rangle$
   & $|12|3|$ & 4 \\
\hline
3 & \ltroisquatre & (120) & \Dtroisquatre
   & $-{*}   o   {*} \langle o\rangle {*} o \rangle$
   & $|1|23|$ & 3 \\
\hline
3 & \ltroiscinq & (111) & \Dtroiscinq
   & ${*} o {*} o {*} o \rangle$
   & $|1|2|3|$ & 1 \\
\hline
\end{tabular}
\caption{Correspondence between several representations of the
RS terms for $n$=1,2,3:
numbered tree, Bloch sequence, Dyck path, bracketing,
non-crossing partition and item in Olszewski's list of
examples \cite{Olszewski-04}.
\label{tab123}}
\end{table}

\begin{table}
\centering
\begin{tabular}{|c|c|c|c|c|c|}\hline
 Tree & Bloch & Dyck & Bracketing & Partition & \cite{Olszewski-04}\\
\hline\hline
 \lquatreun & (4000) & \Dquatreun 
   & $-{*}  \langle o  \rangle {*} \langle o\rangle {*} \langle
    o\rangle  {*} o \rangle$
   & $|1234|$ & 5 \\
\hline
\lquatredeux & (3100) & \Dquatredeux
   & ${*}  \langle o  \rangle {*} \langle o\rangle {*} 
    o  {*} o \rangle$
   & $|134|2|$ & 6 \\
\hline
\lquatretrois & (3010) & \Dquatretrois
   & ${*}  \langle o  \rangle {*} \langle o {*} 
    o\rangle  {*} o \rangle$
   & $|124|3|$ & 14 \\
\hline
\lquatrequatre & (2200) & \Dquatrequatre
   & ${*}  \langle o  \rangle {*}  o {*} \langle
    o\rangle  {*} o \rangle$
   & $|14|23|$ & 7 \\
\hline
\lquatrecinq & (2110) & \Dquatrecinq
   & $-{*}  \langle o  \rangle {*}  o {*} 
    o  {*} o \rangle$
   & $|14|2|3|$ & 4 \\
\hline
\lquatresix & (3001) & \Dquatresix
   & ${*}  \langle o   {*}  o\rangle {*} \langle
    o\rangle  {*} o \rangle$
   & $|123|4|$ & 13 \\
\hline
\lquatresept & (2101) & \Dquatresept
   & $-{*}  \langle o  {*}  o \rangle {*} 
    o  {*} o \rangle$
   & $|13|2|4|$ & 11 \\
\hline
\lquatrehuit & (2020) & \Dquatrehuit
   & ${*}  \langle o   {*} \langle o\rangle {*} 
    o\rangle  {*} o \rangle$
   & $|12|34|$ & 10 \\
\hline
\lquatreneuf & (2011) & \Dquatreneuf
   & $-{*}  \langle o   {*}  o {*} 
    o\rangle  {*} o \rangle$
   & $|12|3|4|$ & 9 \\
\hline
\lquatredix & (1300) & \Dquatredix
   & ${*}   o   {*} \langle o\rangle {*} \langle
    o\rangle  {*} o \rangle$
   & $|1|234|$ & 8 \\
\hline
\lquatreonze & (1210) & \Dquatreonze
   & $-{*}   o  {*} \langle o\rangle {*} 
    o  {*} o \rangle$
   & $|1|24|3|$ & 3 \\
\hline
\lquatredouze & (1201) & \Dquatredouze
   & $-{*}  o  {*} \langle o {*} 
    o\rangle  {*} o \rangle$
   & $|1|23|4|$ & 12 \\
\hline
\lquatretreize & (1120) & \Dquatretreize
   & $-{*}  o  {*} o {*} \langle
    o\rangle  {*} o \rangle$
   & $|1|2|34|$ & 2 \\
\hline
\lquatrequatorze & (1111) & \Dquatrequatorze
   & ${*}   o  {*} o {*} 
    o  {*} o \rangle$
   & $|1|2|3|4|$ & 1 \\
\hline
\end{tabular}
\caption{Correspondence between several representations of the
RS terms for $n$=4:
numbered tree, Bloch sequence, Dyck path, bracketing,
non-crossing partition and item in Olszewski's list of
examples \cite{Olszewski-04}.
\label{tab4}}
\end{table}

\subsection{Bloch sequences}
Bloch~\cite{Bloch-58} was the first to write the general term of $\Omega$
for degenerate systems.
To describe his result, we consider the wave operator
$\Omega(\lambda)$ of the Hamiltonian $H_0+\lambda V$.
We have the series expansion
\begin{eqnarray*}
\Omega(\lambda) = P + \sum_{n=1}^\infty \lambda^n
\Omega_n,
\end{eqnarray*}
with
\begin{eqnarray}
\Omega_n = \sum_{k_1,\dots,k_n} S^{(k_1)} V S^{(k_2)} V \dots V
S^{(k_n)} V P,
\label{defOmegan}
\end{eqnarray}
where $S^0 = - P$, $S^{(k)} = R^k$ for $k>0$,
and where $k_1,\dots,k_n$ run over the Bloch sequences.
A Bloch sequence is an $n$-tuple 
$(k_1 k_2 \dots k_n)$ of non-negative integers, such that
$k_1+\dots+k_m\ge m$ for $m<n$ and
$k_1+\dots+k_n=n$.
The Bloch sequences for $n$=1, 2 and 3 are given
in table~\ref{tab123} and for $n$=4 in table~\ref{tab4}.
These combinatorial objects are enumerated by
Catalan numbers (see Example 6.24 p.~180 of ref.~\cite{Koshy}
and item $m^5$ in Stanley's Catalan addendum).
Since Bloch's publication~\cite{Bloch-58}, they
were widely used~\cite{Messiah-GB,SO,SOE,KuoOsnes} to represent 
the general term of the RS series for degenerate systems. 

The bijection between trees and sequences is defined by
$\phi(\Y) := (1)$, $\phi(s\vee \|):=K\phi(s)$, $\phi(\|\vee s):=(1)\cdot \phi(s)$ and $\phi(s\vee t):=K\phi(s)\cdot\phi(t)$
where $s\not=\|$, $t\not=\|$, where the product
on the right hand sides is the concatenation of sequences
(i.e. $(k_1\dots k_m)(k_{m+1}\dots k_n)=(k_1\dots k_n)$), and
where the operation $K$ acts by
$K(k_1,k_2,\dots,k_n) = (k_1+1,k_2,\dots,k_n,0)$.
Notice that $\phi$ is clearly one to one (injective). The fact that it is actually a bijection follows e.g. from the fact that trees and Bloch sequences are both enumerated by Catalan numbers.

We prove now its compatibility with the tree and Bloch expansions by induction. 
We write $t=t_1\vee t_2$ and we consider the
usual four cases. 
The first case is $t=\Y$, $t_1=t_2=\|$.
This is the starting point of the inductive proof.
We have
$\Omega_t = R V P =  S^1 V P = \Omega_1$,
where we used the fact that the only Bloch sequence
for $n=1$ is $(1)$.
Let us denote by $\Omega_B$ the contribution of $\Omega_n$
in eq.~(\ref{defOmegan}) corresponding to the Bloch sequence
$B=(k_1\dots k_n)$.
Thus, we showed that, for $t=\Y$ and
$B=(1)$, we have $\Omega_t=\Omega_B$. 

Now, assume that $\Omega_t=\Omega_{\phi(t)}$ 
for all trees $t$ with $|t|\le n$ and choose
a tree $t$ such that $|t|=n+1$.
Then, $t=t_1\vee t_2$ with
$|t_1|\le n$ and $|t_2|\le n$.
We consider the three remaining cases.
If $t_1\not=\|$ and $t_2=\|$, then $\Omega_{t}$ is obtained from
eqs.~(\ref{defOmegat}) and (\ref{rect1})
$\Omega_t  = - R \Omega_{t_1} PVP$.
By the induction hypothesis, we 
have $\Omega_{t_1}=\Omega_{B_1}$ for $B_1=\phi(t_1)$.
Therefore, $\Omega_t  = R \Omega_{B_1} (-P)VP$.
If $B_1=(k_1\dots k_n)$, then 
$\Omega_t=\Omega_B$ with
$B=(k_1+1,k_2,\dots ,k_n , 0)=KB_1$ and the property is proved.
The second case is $t_1\not=\|$ and $t_2\not=\|$.
Then eqs.~(\ref{defOmegat}) and (\ref{rectt}) give us
$\Omega_t  = R \Omega_{t_1} (-P)VQ \Omega_{t_2}$.
If $\phi(t_1)=B_1=(k_1\dots k_p)$ and
$\phi(t_2)=B_2=(l_1\dots l_q)$, then
$\Omega_t=\Omega_B$ with
$B=(k_1+1,k_2,\dots ,k_p, 0, l_1,\dots ,l_q)=K(B_1) B_2$.
Thus, $\phi(t_1\vee t_2)=K(B_1) B_2=K\phi(l_1)\cdot\phi(l_2)$.
The last case is $t_1=\|$ and $t_2\not=\|$. 
Equations~(\ref{defOmegat}) and (\ref{rec1t})
give us $\Omega_t  = R VQ \Omega_{t_2}$.
If $B_2=\phi(t_2)=(k_1\dots k_n)$, then
$\Omega_t=\Omega_B$ with
$B=(1k_1\dots k_n)=(1)\cdot\phi(t_2)$.

\subsection{Dyck paths}
In ref.~\cite{Bloch-58}, Bloch also defined geometrical
objects that are Dyck paths rotated by $\pi/4$.
We prefer to use Dyck paths because they are 
thoroughly studied in the combinatorial literature
(see, for example
Item i, p.~221 of ref.~\cite{StanleyI} or
Example 6.2, p.~151 of ref~\cite{Koshy}, where they are
called \emph{mountain ranges}).

The bijection between Dyck paths and Bloch sequences
used in ref.~\cite{Bloch-58} is well known in the
combinatorial literature~\cite[p.~168 and 181]{Koshy}.
To build the Dyck path corresponding to the the Bloch sequence 
$(k_1 k_2 \dots k_n)$, start from
the origin and make $k_1$ steps in the North-East direction,
then make one step in the South-East direction, 
then $k_2$  steps in the North-East direction,
then  one step in the South-East direction, 
and so on.
This bijection is illustrated in tables~\ref{tab123} and \ref{tab4}.

\subsection{Non-crossing partitions}
The terms of the RS series of degenerate systems
can also be described by non-crossing partitions.
This correspondence was studied by Olszewski~\cite{Olszewski-04} because it
leads to useful factorizations of the RS terms for non-degenerate systems.

Consider the examples at the beginning of 
section~\ref{bijection-sect}.
If we just look at the denominators, we see that each
of them can be deduced from the right comb by saying that 
some indices are equal.
For the five trees with $|t|=3$, the term corresponding to
$t=\troisun$ can be obtained from that of $t=\troiscinq$ by 
stating that $e_{i_1}=e_{i_2}=e_{i_3}$.
The other terms follow from
$e_{i_1}=e_{i_3}$,
$e_{i_1}=e_{i_2}$ and
$e_{i_2}=e_{i_3}$.

More generally, for
a given tree $t$, we say that two indices $j$ and $k$
are equivalent if and only if $e_{i_j}=e_{i_k}$ in the
denominator of $\Omega_t$.
The sets of equivalent indices form a partition of $\{1,\dots,n\}$.
We recall that a partition $B_1|B_2|\dots|B_k$ of $\{1,\dots, n\}$
is a set of disjoint subsets $B_i$ of $\{1,\dots, n\}$
whose union is $\{1,\dots, n\}$. Each subset $B_i$
is called a block of the partition.
The partitions corresponding to the RS terms of order 1 
to 3 are given in table~\ref{tab123} and for order 4 in table~\ref{tab4}.
It will be shown that these partitions are \emph{non-crossing}.
Two blocks $A$ and $B$ of a partition are said to be crossing
if there are $a,b$ in $A$ and $x,y$ in $B$ such that
$a<x<b<y$ or $x<a<y<b$. A partition is called non-crossing if
no two of its blocks cross.

We build by induction a partition from a tree.
Note first that for tree $t$ of order $n$ we deal
with partitions of $\{1,\dots,n\}$. A tree of
order $n$ has $n+1$ leaves. Thus, the index of the rightmost
leaf is not used in the partition.

For a given tree $t=t_1\vee t_2$, we call
$P$ ($P_1$, $P_2$, respectively) the partition
corresponding to $t$ ($t_1$, $t_2$, respectively).
We consider the usual four cases.
(i) For $t_1=t_2=\|$ we associate the partition $P=|1|$.
(ii) For $t_1\not=\|$ and $t_2=\|$,
eq.~(\ref{rect1}) gives us the additional
denominator $e_0-e_1$. Therefore, an additional index is equivalent
to 1. This index is that of the rightmost leaf of $t_1$, which
was not used in $P_1$. Therefore, the partition
$P$ of $t$ is obtained from $P_1$ by adding to the block of $P_1$
containing 1 the index of the rightmost leaf of $t_1$,
which is $|t_1|+1$.
(iii) For $t_1=\|$ and $t_2\not=\|$, 
eq.~(\ref{rec1t}) gives us the additional
denominator $e_0-e_1$. However, the leaf denoted by 1 is new
and no block of $P_2$ should contain it.
Therefore, the partition
$P$ of $t$ is obtained from $P_2$ by increasing all
numbers of $P_2$ by 1 and by
adding the block $|1|$. 
(iv) For $t_1\not=\|$ and $t_2\not=\|$, we compose the previous
cases. We build $P$ by adding the index $|t_1|+1$ to the
block of $P_1$ containing the index 1 and we increase all
indices of $P_2$ by $|t_1|+1$.
We check that, in all cases, the resulting partition is non-crossing.

Note that Olszewski~\cite{Olszewski-04} does not explicitly use
non-crossing partitions. He describes the general term of the RS 
series for non-degenerate systems by drawing a circle with $n$ 
points and pinching some of these points. The relation
with non-crossing partitions is straightforward: all
points that are pinched together belong to the same block
of the partition. However, his correspondence between partitions and
terms of the RS series is not the same as ours.

\subsection{Bracketing}
Following a suggestion by Brueckner~\cite{Brueckner-55},
Huby and Tong~\cite{Huby-61,Tong-62} proposed, for the energy
of nondegenerate systems, a solution in terms of bracketing, that is
isomorphic with Stanley's problem $e^5$ in his
``Catalan addendum''.

Consider the example of $t=\deuxun$. We have
$\Omega_t=R^2 VPVP$. For a nondegenerate system
$P=|0\rangle\langle 0|$ and
$\Omega_T|0\rangle = R^2 V|0\rangle \langle 0|V|0\rangle$.
The rule of the game is now to insert 
expectation values to disjoin powers of $R$. Thus,
$\Omega_T|0\rangle = -R\langle 0|V|0\rangle R V|0\rangle$.
We use Tong's pictorial representation, where
$R$ is replaced by $*$, $V$ by $o$,
$\langle 0|$ by $\langle$ and 
$|0\rangle$ by $\rangle$, so that
$\Omega_T|0\rangle = - * \langle o \rangle * o\rangle$.
The bracketings for $|t|$=1,2 and 3 are given in 
table~\ref{tab123} and for order 4 in table~\ref{tab4}.
Brueckner's bracketing is a powerful way to simplify
the RS series, for instance by including the
vacuum expectation value $\langle 0| V |0\rangle$
into $H_0$, so that all RS terms involving it cancel.
More details and examples can be found in some
textbooks~\cite{Shavitt} or review papers~\cite{Paldus-75}.

We now describe the connection between trees and bracketings.
Denote by $b(t)$ the bracketing corresponding to $t$
in Tong's representation. Assume that $b(t)$ is known
for all trees $t$ with $|t|\le n$ and take a
tree $t=t_1\vee t_2$ of degree $n$.
It $t_1=\|$ and $t_2\not=\|$, then
$\Omega_t|0\rangle =  RV \Omega_{t_2}|0\rangle$,
so that $b(t)={*}o b(t_2)$.
It $t_1\not=\|$ and $t_2=\|$, then
$\Omega_t|0\rangle =  -R \Omega_{t_1}|0\rangle\langle 0|V|0\rangle
 =  -R \langle 0|V|0\rangle \Omega_{t_1}|0\rangle$,
so that $b(t)=-{*}\langle o\rangle  b(t_1)$.
It $t_1\not=\|$ and $t_2\not=\|$, then
$\Omega_t|0\rangle = -R\Omega_{t_1}|0\rangle
\langle 0| V \Omega_{t_2}|0\rangle=
 -R\langle 0| V \Omega_{t_2}|0\rangle \Omega_{t_1}|0\rangle$,
so that $b(t)=-{*}\langle o b(t_2) b(t_1)$.
Note that this bijection is different from the one 
used by Tong~\cite{Tong-62}.

An equivalent bijection is obtained by numbering
the inner vertices of the trees.
The operation $\nu$ that associates to each tree $t$ its numbered tree
is defined as follows.
We denote by $\nu(t)[k]$ the numbered tree
obtained from $\nu(t)$ by adding $k$ to all the vertex numbers.
Then, $\nu(t)$ can be defined recursively.
If $t=\|$, then $\nu(t)=\|$ (no number).
If $t=\Y$, then $\nu(t)$ is obtained by assigning the number 1
to the root.
If $t=t_1\vee  t_2$, then the root has number 1,
the inner vertices of $t_2$ (if any) are numbered as $\nu(t_2)[1]$ and
the inner vertices of $t_1$ (if any) are numbered as $\nu(t_1)[|t_2|+1]$.
The numbered trees for $|t|\le 3$ are given in table
table~\ref{tab123} and for $|t|=4$ in table~\ref{tab4}.
For any tree $t$, we can build the sets of numbers
belonging to the same line oriented to the left.
For example, if $t=\,\,\,\quatredouze$, the numbering is
$\,\,\,\lquatredouze$ and the sets are $|1|24|3|$.
These sets form a non-crossing partition which is the
same as that used by Olszewski~\cite{Olszewski-04}.
To obtain $b(t)$, first number the $|t|$ stars of 
$b(t)$ from 1 to $|t|$ from the left to the right.
Then, consider a block $B=k_1\dots k_p$
of the partition. If $p=1$, do nothing,
if $p>1$, then write a $\langle$ after star number $k_1$, a
$\rangle$ before star number $k_p$ and 
replace star number $k_i$ with $1<i<p$ (if any)
by $\rangle * \langle$.

\section{Conclusion}
Combinatorial physics is an emerging field that uses
modern tools of algebraic combinatorics to solve
physical problems.
It was born with the investigation of the algebraic
structure of renormalization in quantum field 
theory~\cite{CKI} and showed its ability
to deal with many-body problems~\cite{BrouderMN}.

We showed that combinatorial physics is able to solve
such long-standing problems as time-independent
perturbation theory. The RS series is at the heart
of many applications of quantum mechanics.
It is also equivalent to more sophisticated
methods such as Feynman diagrams~\cite{Wu-51}.
It is even related to
Wilson's renormalization group~\cite{Muller-96}.

Our combinatorial methods provided easy resummations
of the RS series. It remains now to test their convergence
properties.

Note that Arnol'd also used trees in perturbation
theory~\cite{Arnold-83}. However, his trees are essentially
different from ours because they describe the successive
degeneracy splitting due to higher order terms~\cite{Tsaune}.

\section{Acknowledgement}
G. S. T. was partially supported by the grant OTKA K60040.


\begin{thebibliography}{10}
\bibitem{suzuki-quasi-degenerate}
This problem was later solved by Suzuki and coll.~\cite{Suzuki-94}.
\end{thebibliography}

\end{document}